\documentclass[structabstract]{aa}

\usepackage{graphicx}
\usepackage{natbib}
\usepackage{amsmath}
\usepackage{amssymb}
\usepackage{eurosym}
\usepackage{latexsym}
\usepackage{subfig}
\usepackage{color}

\begin{document}

   \title{The role of filament activation in a solar eruption}

   \author{F.~Rubio~da~Costa\thanks{Now at Max-Planck-Institut f\"{u}r Sonnensystemforschung. Max-Planck-Strasse 2, 37191 Katlenburg-Lindau, Germany}
          \inst{1,2}
          \and
          F.~Zuccarello
          \inst{1}
          \and
          L.~Fletcher
          \inst{2}
	  \and
	  P. Romano
          \inst{3}
	  \and
	  N. Labrosse
          \inst{2}}

   \institute{Department of Physics and Astronomy,
              University of Catania,
              Via S. Sofia 78, 95123,
              Catania, Italy.\\
              \email{rubio@mps.mpg.de}
         \and
              School of Physics and Astronomy,
              SUPA,
              University of Glasgow,
              Glasgow, G12 8QQ,
              U. K.\\
         \and
	      INAF-Catania Astrophysical Observatory,
              Via S. Sofia 78, 95123,
              Catania, Italy.\\
}

   \date{Received 15 March 2011 / Accepted 16 December 2011}

\newcommand{\lya}{Ly$\alpha$}
\newcommand{\ha}{H$\alpha$} 
 
  \abstract
   {Observations show that the mutual relationship between filament eruptions and solar flares cannot be described in terms of an unique scenario. In some cases, the eruption of a filament appears to trigger a flare, while in others the observations are more consistent with magnetic reconnection that produces both the flare observational signatures (e.g., ribbons, plasma jets, post-flare loops, etc.) and later the destabilization and eruption of a filament.}
   {Contributing to a better comprehension of the role played by filament eruptions in solar flares, we study an event which occurred in NOAA 8471, where a flare and the activation of (at least) two filaments were observed on 28 February 1999.}
   {By using imaging data acquired in the 1216, 1600, 171 and 195~\AA~TRACE channels and by BBSO in the continnum and in the \ha\ line, a morphological study of the event is carried out. Moreover, using TRACE 1216 and 1600~\AA\ data, an estimate of the ``pure'' \lya~ power is obtained. The extrapolation of the magnetic field lines is done using the SOHO/MDI magnetograms and assuming a potential field.}
  {Initially an area hosting a filament located over a $\delta$ spot becomes brighter than the surroundings, both in the chromosphere and in the corona. This area increases in brightness and extension, eventually assuming a two-ribbon morphology, until it reaches the eastern part of the active region. Here a second filament becomes activated and the brightening propagates to the south, passing over a large supergranular cell. The potential magnetic field extrapolation indicates that the field line connectivity changes after the flare.}
    {The event is triggered by the destabilization of a filament located between the two polarities of a $\delta$ spot. This destabilization involves the magnetic arcades of the active region and causes the eruption of a second filament, that gives rise to a CME and to plasma motions over a supergranular cell. We conclude that in this event the two filaments play an active and decisive role, albeit in different stages of the phenomenon, in fact the destabilization of one filament causes brightenings, reconnection and ribbons, while the second one, whose eruption is caused by the field reconfiguration resulting from the previous reconnection, undergoes the greatest changes and causes the CME.}


   \maketitle

\section{Introduction}
Observations show that some phenomena occurring in the solar atmosphere, such as filament eruptions, flares and coronal mass ejections (CMEs) are often  related to each other. Recently, \citet{2004ApJ...614.1054J} presented a statistical study of filament eruptions, showing that about 55 \% of the 98 events analyzed were associated with CMEs and that active region filament eruptions have a considerably higher flare association rate (95 \%) compared to the eruption of quiescent filaments (27 \%). In general, the time sequence indicates that firstly a filament is activated and starts to rise and within some tens of minutes, the flare and/or the CME occurs (see, e.g., \citealt{1976SoPh...50...85K}). Phenomena commonly associated with these energetic processes are filaments rising, particle acceleration, bright ribbons in the lower atmospheric layers and post-flare loops.

The study of these phenomena has indicated that the triggering mechanism is related to an unstable magnetic configuration. More specifically, the magnetic field can become stressed (i.e., non-potential) and therefore can increase its energy when shearing, converging or twisting motions of the footpoints of its field lines take place \citep{1994JGR....9921467P, 2001ApJ...551..586H, 1999ApJ...510..485A}. In addition, an unstable magnetic field configuration can be reached due to new magnetic flux emergence and interaction with pre-existing magnetic field \citep{1977ApJ...216..123H}. One possible consequence of these processes is the occurrence of magnetic reconnection, which causes the re-arrangement of the magnetic configuration, and the conversion of the stored magnetic energy into thermal, radiative and kinetic energy (see e.g., \citealt{2000mare.book.....P, 2002A&ARv..10..313P, 2004psci.book.....A}).

Observations indicate that the temporal sequence of these phenomena, and especially the role played by filament activation and eruption can be significantly different from event to event \citep{2005ApJ...630.1148S, 2007AAS...210.9321W, 2009ApJ...703..757L, 2009A&A...493..629Z}.  Multi-wavelength observations allow us to investigate the behavior of features at different atmospheric levels, which helps address whether and in what conditions the eruption of the filament plays an \textit{active} role (the filament eruption destabilizes the magnetic field, giving rise to the flare emission) or a \textit{passive} role (the filament eruption follows in time the flare emission) in the flare/CME occurrence. 

The study of filament destabilization and eruption can be carried with chromospheric data in the H$\alpha$ line and, as has been recently highlighted, also with observations in the TRACE 1600 and 1216 \AA\ channels. In fact, from 1216 \AA\ observations it is possible to follow the evolution of a filament in the high chromosphere and eventually compare it with what can be observed at a lower atmospheric layer. A study of a solar flare and filament eruption, carried out using TRACE data at 1216, 1600 and 171 \AA, together with Yohkoh hard and soft X-ray data, allowed us, in a previous paper \citep{2009A&A...507.1005R}, to deduce that during the early stage of the filament eruption a roughly circular spreading sheet-like ejecta was produced and that the compact footpoints observed in  \lya\ were well correlated with HXR footpoints.  Also important is the photospheric magnetic field configuration, providing information on the event's magnetic environment. When the active region is located approximately at disk centre,  line-of-sight magnetograms can be used to infer the potential or linear force-free magnetic field configuration and to deduce the morphology of the magnetic field also at higher atmospheric levels - useful for comparison with coronal observations. 

In order to contribute to understanding the role played by filament eruption in the triggering of solar flares,  we study an M6.6 flare which occurred on 28 February 1999 in active region NOAA 8471. The active region was observed by Big Bear Solar Observatory (BBSO) in the continuum and in the \ha\ line, by TRACE at different wavelengths (1216~\AA, 1600~\AA, 171~\AA\ and 195~\AA), and by MDI on board SOHO. In Section \ref{sect:data_analysis} we review the observations and describe the data processing for the event, and in Section \ref{sect:overview} we give an overview of the flare event. The study of the magnetic field configuration at the flare site is presented in Section \ref{sect:magnetic_field} and in Section \ref{sect:disc} we present our discussion and conclusions.

\section{Data Analysis}\label{sect:data_analysis}
\begin{figure}
\centering
  \includegraphics[width=9.cm]{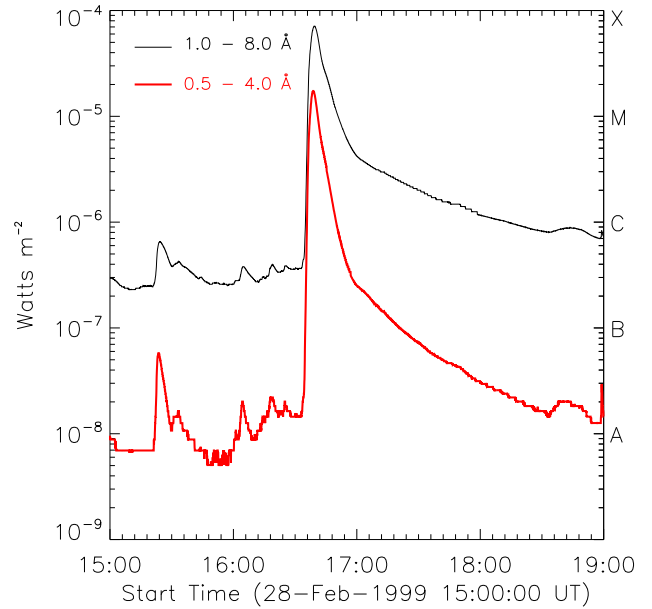}
     \caption{GOES 10 X-ray flux as a function of time measured on 28 February 1999. The data were taken every three seconds.}
     \label{goes_flare2}
\end{figure}

In active region NOAA 8471, at N29 W12, an M6.6 GOES class flare occurred on 28 February 1999, starting at 16:31:00 UT and reaching its peak at 16:38:35 UT. Fig. \ref{goes_flare2} shows the X-ray flux measured by the GOES satellite: after the flare peak there is an initial rapid decrease and later a much smoother gradual phase, lasting about 40 minutes. A coronal mass ejection (CME) was observed by LASCO/SOHO C2 coronograph at 17:54:05 UT. The corresponding time of the CME initiation calculated with the height-time measurements is 16:42  UT (see http://cdaw.gsfc.nasa.gov/CME\_list/) and taking into account the location and the timing, we can associate the CME with the M6.6 class flare.

The data used to investigate this event are as follows:

TRACE data are acquired at 1216, 1600, 171 and 195~\AA\ (see Table \ref{table_trace_data2}). The data are available till the maximum of the flare (pre-flare and impulsive phase), but there are no data for the gradual decay phase.

\begin{table}
\centering
\caption{TRACE data available for the 28 Feb. 1999 flare.}
\begin{tabular}{c c }
\hline\hline
Filter & Time Range (UT) \\
\hline
1216 & 12:01:38 - 16:41:15 \\
1600 & 15:00:54 - 16:41:08 \\
171 & 16:10:54 - 16:41:32 \\
195 & 13:00:09 - 16:41:42 \\
\hline
\end{tabular}
\label{table_trace_data2}
\end{table}

The raw images acquired by TRACE were corrected for instrumental effects (subtracting the dark current and pedestal, correcting for exposure time, for radiation spikes or for saturated pixels) using the standard procedures included in the Solar Software IDL routines provided by the TRACE team \citep[SSW;][]{1998SoPh..182..497F}. We also corrected for solar rotation, and for offsets between the 1216~\AA\ and 1600~\AA\ channels by measuring the displacement of network bright points in images acquired close in time. The offset between the 1216~\AA\ and the 1600~\AA\ images is -2.19 pixels in x and 1.24 pixels in y.

\ha\ full disk images were acquired with the 25 cm telescope at the Big Bear Solar Observatory during the time interval 15:31:04 - 18:34:24 UT and 22:09:13 - 23:51:29 UT. The pixel size of the images is of 1.07 arcsec/pixel. The images were taken using a H$\alpha$ filter, centered at the center of the \ha\ line (6562.8~\AA) and with a bandpass of 0.25~\AA.

\ha\ (512 $\times$ 512 pixels) images were acquired by the 65 cm telescope at the Big Bear Solar Observatory, with a pixel size of 1.158 arcsec/pixel. Further images are available for the previous day and the day after the flare (See Table \ref{table_ha_data}). All the images were rotated by the P-angle. The images were taken using a \ha\ filter, centered at the center of the \ha\ line (6562.8~\AA) and with a bandpass of 0.25~\AA. For the 27 February there are also images acquired along the \ha\ profile at $\pm$0.25, $\pm$0.5, $\pm$0.75 and $\pm$1~\AA.

\begin{table}[!h]
\centering
\caption{BBSO H$\alpha$ data available for NOAA 8471.}
\begin{tabular}{c c }
\hline\hline
Day & Time Range (UT)\\
\hline
27-02-99 & 18:52:25 - 21:03:32 \& 23:00:06 - 23:49:51 \\
28-02-99 & 17:48:19 - 21:36:03 \& 23:02:29 - 23:30:58 \\
01-03-99 & 18:17:07 - 18:19:05 \\
\hline
\multicolumn{2}{l}{Note: H$\alpha$ data (1.158 arcsec/pixel) from }\\
\multicolumn{2}{l}{27 Feb. to 01 March 1999.}\\
\end{tabular}
\label{table_ha_data}
\end{table}

High resolution (512 $\times$ 481 pixels) images of the 4 components of the Stokes parameters at 6103~\AA\ were acquired with the 25 cm telescope at the Big Bear Solar Observatory, with a pixel size of 0.683 arcsec/pixel. There are also some data available for 27 February and 01 March 1999 (See Table \ref{table_stockes_data}).

\begin{table}[!h]
\caption{BBSO Stokes data available for NOAA 8471.}
\centering
\begin{tabular}{c c c}
\hline\hline
Day & Parameter & Time Range (UT)\\
\hline
27-02-99 & I & 16:55:17 - 23:49:45 \\
& V & 16:54:56 - 23:49:23\\
28-02-99 & I & 16:01:23 - 23:30:52 \\
& V & 16:01:01 - 23:36:38 \\
01-03-99 & I & 17:30:30 - 01:06:03 \\
& V & 17:30:08 - 01:05:14 \\
\hline
\multicolumn{3}{l}{Note: Stokes I and V data from 27 Feb.}\\
\multicolumn{3}{l}{to 01 March 1999.}\\
\end{tabular}
\label{table_stockes_data}
\end{table}

The Stokes-I component is interpreted as a photospheric continuum image. The ratio of the Stokes-V component to the Stokes-I component provides a line-of-sight magnetogram \citep{1992soti.book...71L}. We used high resolution BBSO data, and in particular the Stokes I and V components in order to obtain detailed photospheric images and maps of the magnetic configuration, complementing the MDI data when their spatial resolution was not sufficient for our purposes.

Full disk MDI/SOHO magnetograms at Ni I 6767.8~\AA, were taken on 27 and 28 February 1999, which provide the component of the magnetic field along the line of sight. The MDI/SOHO full disk magnetograms are 1024 $\times$ 1024 pixels image with a pixel size of 1.98 arcsec/pixel and a time resolution of 96 minutes.

We deduced the pixel size and the field of view of the data provided by BBSO using the following method: we compared the high resolution images of the Stokes V component with the full disk MDI/SOHO magnetograms by overlapping the contours corresponding to the penumbral edges of the main sunspots of the active region. From the best fit of the overlapping contours we determined the pixel size of all the images of the 4 components of the Stokes parameters. Then we compared the images of the Stokes I component with the high resolution \ha\ images taken at + 1.0~\AA\ from the center of the line. In this case the best fit of the contours of the umbra and penumbra of the main sunspots provided us the pixel scale of images taken by the 65 cm BBSO telescope. Finally, by the comparison of the high resolution and the full disk \ha\ images we deduced the pixel scale for the H$\alpha$ images taken by the 25 cm telescope.

\section{Overview of the event}\label{sect:overview}

\subsection{Morphology of the active region}
The active region NOAA 8471 appeared on 23 February 1999 at the north-east limb of the Sun as a small beta region with arch-type filaments, bright \ha\ increasing plage and small sunspots. The active region was observed for 10 days and it had associated several flare events (BBSO Solar Report).

\begin{figure*}
\centering
  \includegraphics[width=18.cm]{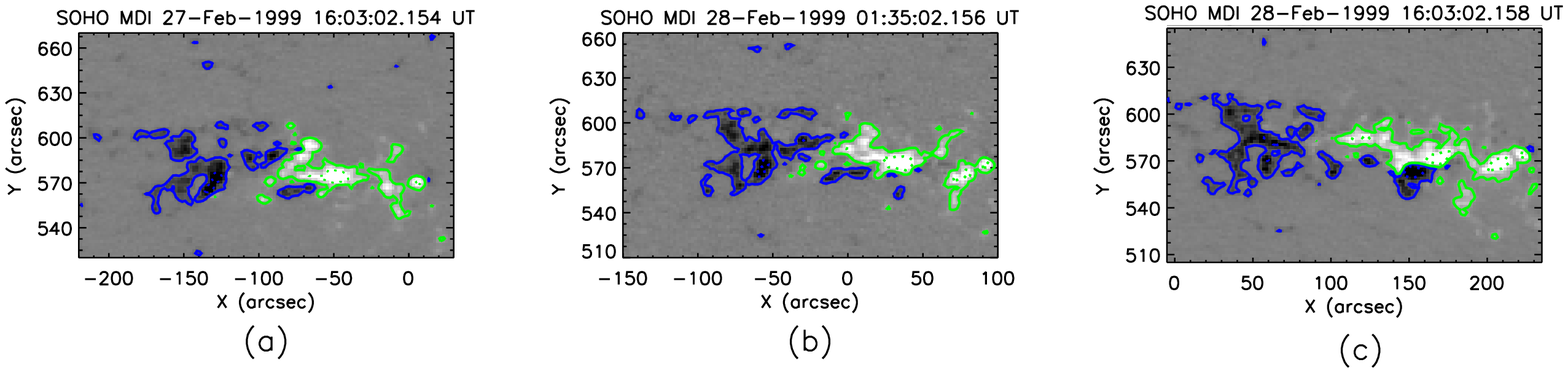}
  \includegraphics[width=18.cm]{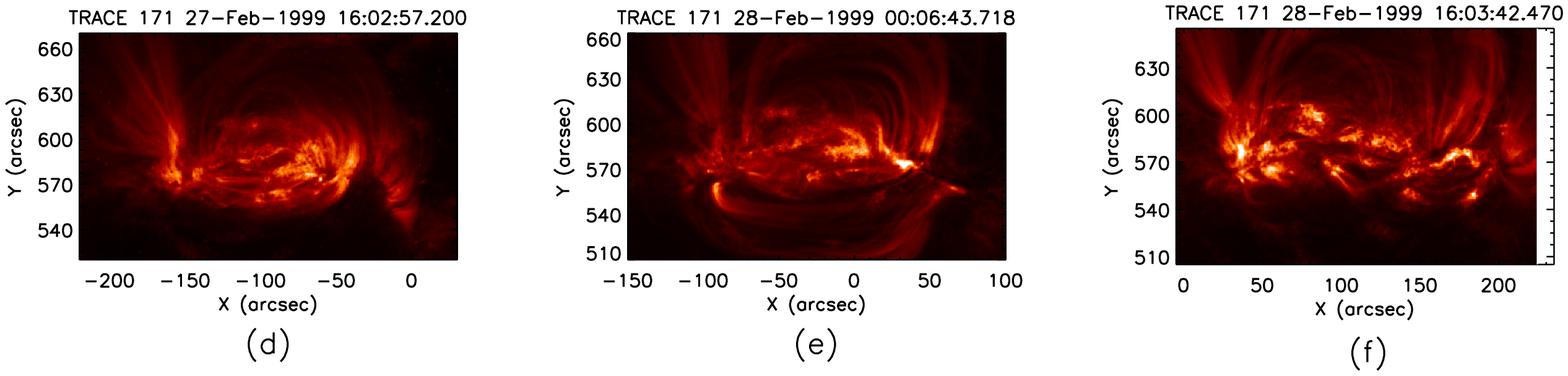}
     \caption{Top row: SOHO/MDI images acquired at (a): 16:03 UT on 27 Feb.; (b): at 01:35 UT on 28 Feb.; (c): at 16:03 UT on 28 Feb. The blue and green contours indicate the negative and positive magnetic field at $\pm$ 200 (solid line), 1000 G (dashed line). Bottom row:  TRACE 171~\AA\ images acquired at (d): 16:03 UT on 27 Feb.; (e): 00:07 UT on 28 Feb.; (f):  16:03 UT on 28 Feb. The images have a field of view of 192 $\times$ 115 Mm$^2$. In this and in the following images North is at the top, West is at the right.}
     \label{evol_mdi_trace171}
\end{figure*}

\begin{figure}[!th]
\centering
  \includegraphics[width=8.cm]{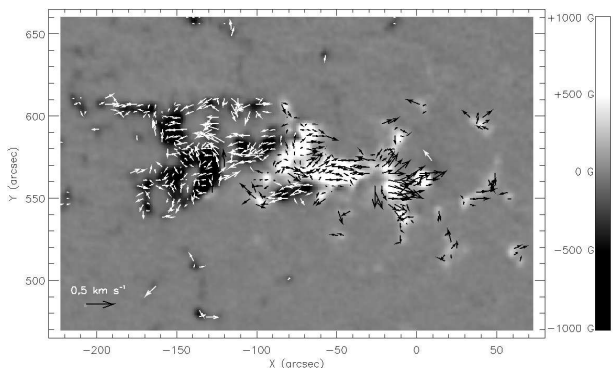}
     \caption{\bf SOHO/MDI line-of-sight magnetogram in active region NOAA 8471 on Feb 27 at 16:52 UT. The arrows represent the transverse velocity field. The field of view is 217 $\times$ 145 Mm$^2$.}
     \label{horiz_veloc}
\end{figure}

In Fig. \ref{evol_mdi_trace171}  a sequence of MDI magnetograms  and TRACE 171~\AA\ images shows the rapid evolution of the active region over the 24 hours starting 16:03 UT on 27 February. In particular, comparing Fig. \ref{evol_mdi_trace171}(a) with Fig. \ref{evol_mdi_trace171}(b), we can see that some knots of negative polarity in the central part of the active region shift towards the east. We measured the horizontal photospheric velocities in MDI magnetograms using the Differential Affine Velocity Estimator (DAVE) \citep{2005ApJ...632L..53S}. We considered magnetogram subfields of $300 \times 200$ arcsec centered in NOAA 8471 and aligned all subfields by applying a standard differential rotation rate \citep{1990SoPh..130..295H} with a sampling of 1 arcsec, i.e., implementing a subpixelization. We corrected all the magnetograms for the angle between the magnetic field direction and the line-of-sight. We used a full-width-at-half maximum of the apodization windows of 19.80 arcsec and a time interval of 96 min.  Fig. \ref{horiz_veloc} shows the SOHO/MDI line-of-sight magnetogram overplotted with the \begin{bf}transverse\end{bf} velocity field, represented by the arrows. Using the DAVE method we deduced that the eastward motion of the knots of negative polarity persists until the end of our dataset, with an average velocity of about 0.2 km s$^{-1}$. This behavior, in addition to the prevalent westward motion of the main positive polarity, indicates the presence of shearing motions in the central part of the active region. Moreover, comparing Fig. \ref{evol_mdi_trace171}(d) with Fig. \ref{evol_mdi_trace171}(e), we see that the morphology of the active region in the corona changed significantly during the same time interval.

The comparison of the MDI magnetograms acquired on 28 February at 01:35 UT and 16:03 UT (Figs. \ref{evol_mdi_trace171} (b) and \ref{evol_mdi_trace171}(c)) shows that while the shearing motion continues in the central part of NOAA 8471,  an extended region of negative polarity appears at position [150, 560] arcsec, which will give rise to a $\delta$ spot (see below). Comparing the TRACE 171~\AA\ images acquired on 28 February at 00:06 UT and 16:03 UT (Figs. \ref{evol_mdi_trace171}(e) and \ref{evol_mdi_trace171}(f)) we can see that during this time interval some relatively small bright loops appeared in the central region (where the shearing motions took place), while the larger loops visible in Fig. \ref{evol_mdi_trace171}(e) in the southern part of the active region are no longer visible in Fig. \ref{evol_mdi_trace171}(f). In the final image, the region has a very elongated magnetic polarity inversion line.

\begin{figure*}[!ht]
\centering
  \subfloat[][Continuum]{\includegraphics[width=6.1cm]{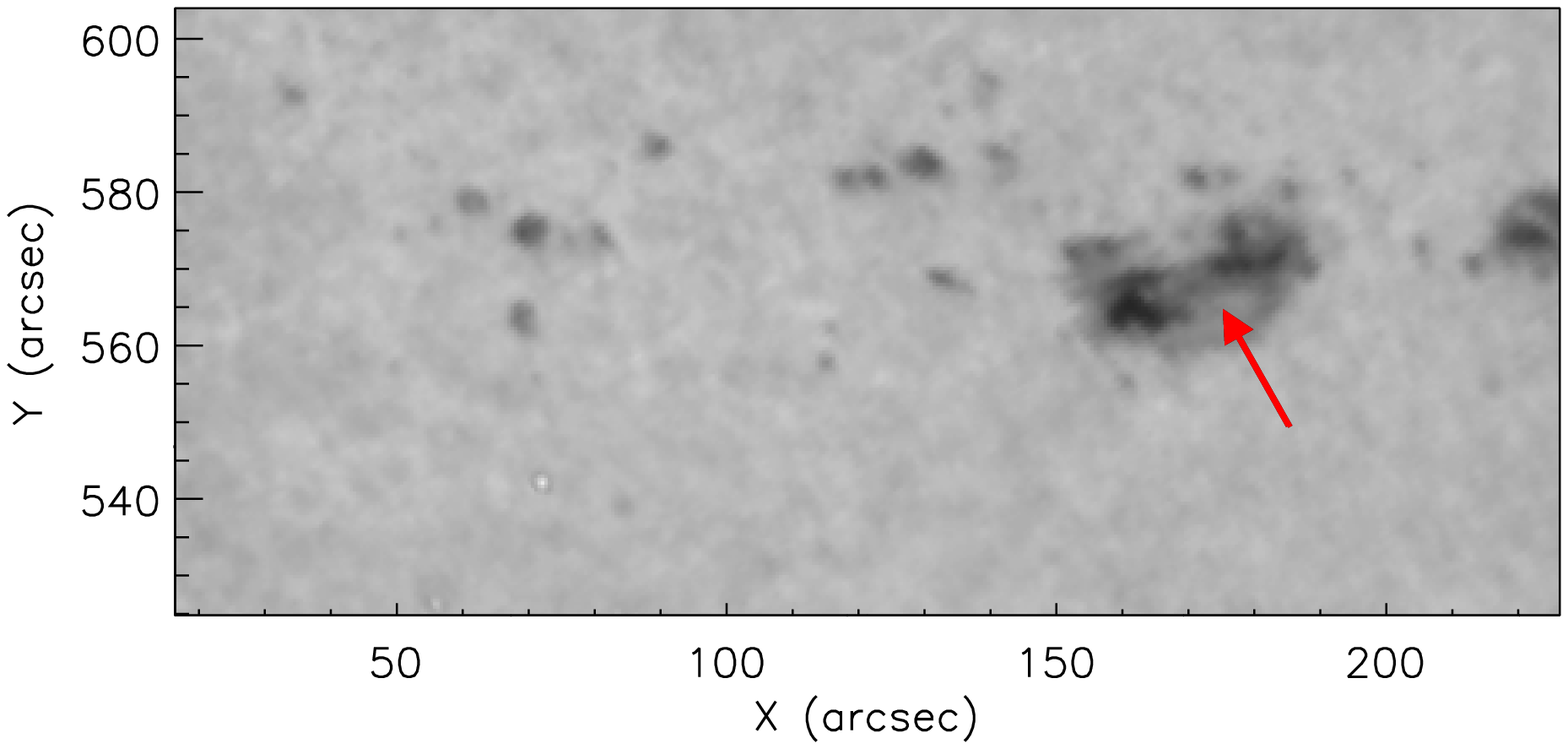}}
  \subfloat[][H$\alpha$]{\includegraphics[width=6.1cm]{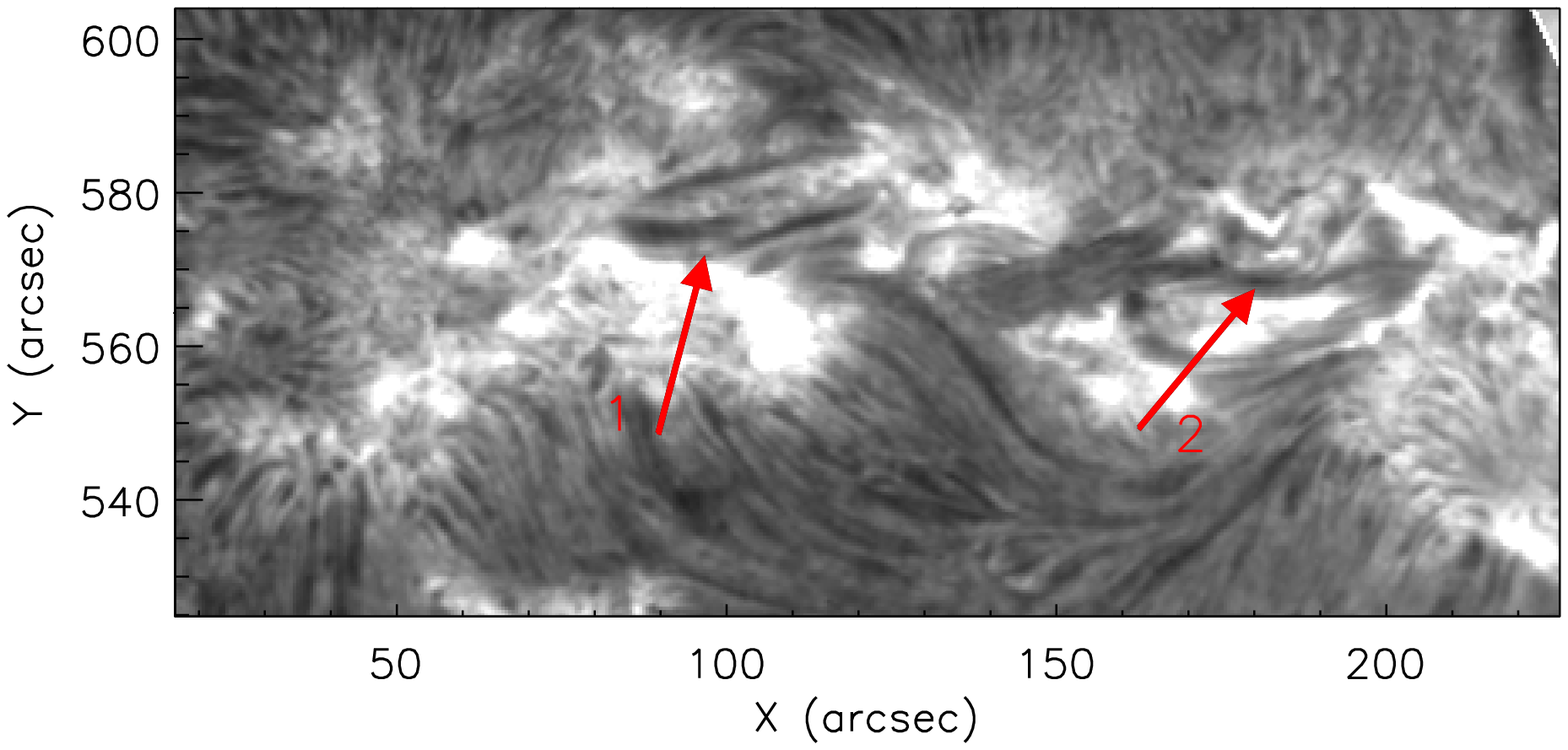}}
  \subfloat[][Magnetogram]{\includegraphics[width=6.1cm]{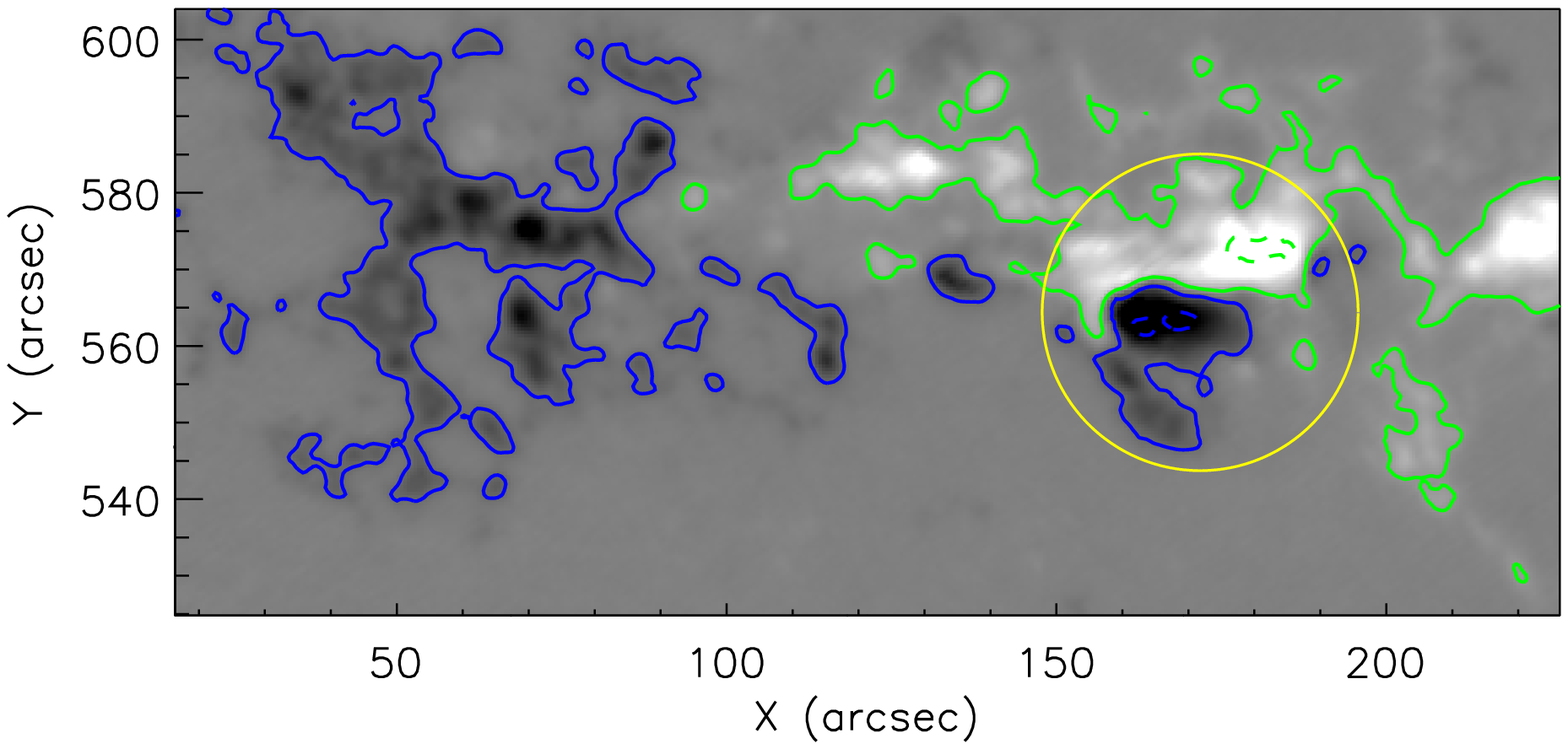}}
     \caption{(a): Image in the continuum acquired at BBSO at 17:49:52 UT on 28 February 1999, the arrow indicates a sunspot with a $\delta$-configuration; (b): High resolution $H_{\alpha}$ image acquired at BBSO at 17:49:48 UT, showing several filaments; the arrows indicate two filaments involved in the flaring process; (c): Magnetogram acquired at BBSO, showing the magnetic configuration of the active region at 17:49:52 UT; the blue and green contours indicate the negative and positive magnetic field at $\pm$200 (solid line), $\pm$1000 G (dashed line). The circle indicates the location of the $\delta$ spot. The images have a field of view of 161 $\times$ 61 Mm$^2$.}
     \label{config_flare2}
\end{figure*}

The continuum image in Fig. \ref{config_flare2}(a) from BBSO on 28 February 1999 at 17:49:52 UT shows a sunspot-group characterized by a $\beta \delta$ magnetic configuration. On the western side is the main sunspot with the $\delta$ configuration (indicated by an arrow) and another sunspot with positive polarity; some pores, with both positive and negative polarities, are present in the central and eastern side (compare with Fig. \ref{config_flare2}(c)).

The \ha\ image (Fig. \ref{config_flare2}(b)), taken at BBSO on 28 February 1999 at 17:49:48 UT, shows a bright facular region, and several filaments crossing the active region from east to west. TRACE 171 and 195~\AA\ images indicate the corresponding EUV filament channels (arrows in Fig. \ref{TRACE_flare2} (c)-(d)). 
In Fig. \ref{config_flare2}(c), the BBSO high resolution line-of-sight magnetogram, taken on 28 February 1999 at 17:50:31 UT, shows the magnetic configuration of the active region, which appears quite simple, with a clear separation between the opposite polarities. Comparing both continuum and H$\alpha$ images with the magnetogram, obtained with the Stokes parameters V/I, we can see that the main sunspot, characterized by $\delta$ configuration, is situated over the circled region in Fig. \ref{config_flare2}(c), while the brightest facular region visible in the eastern part of the \ha\ image corresponds to the negative polarity of the active region.

\begin{figure*}
\centering
  \subfloat[][27-02-1999]{\includegraphics[width=6.1cm]{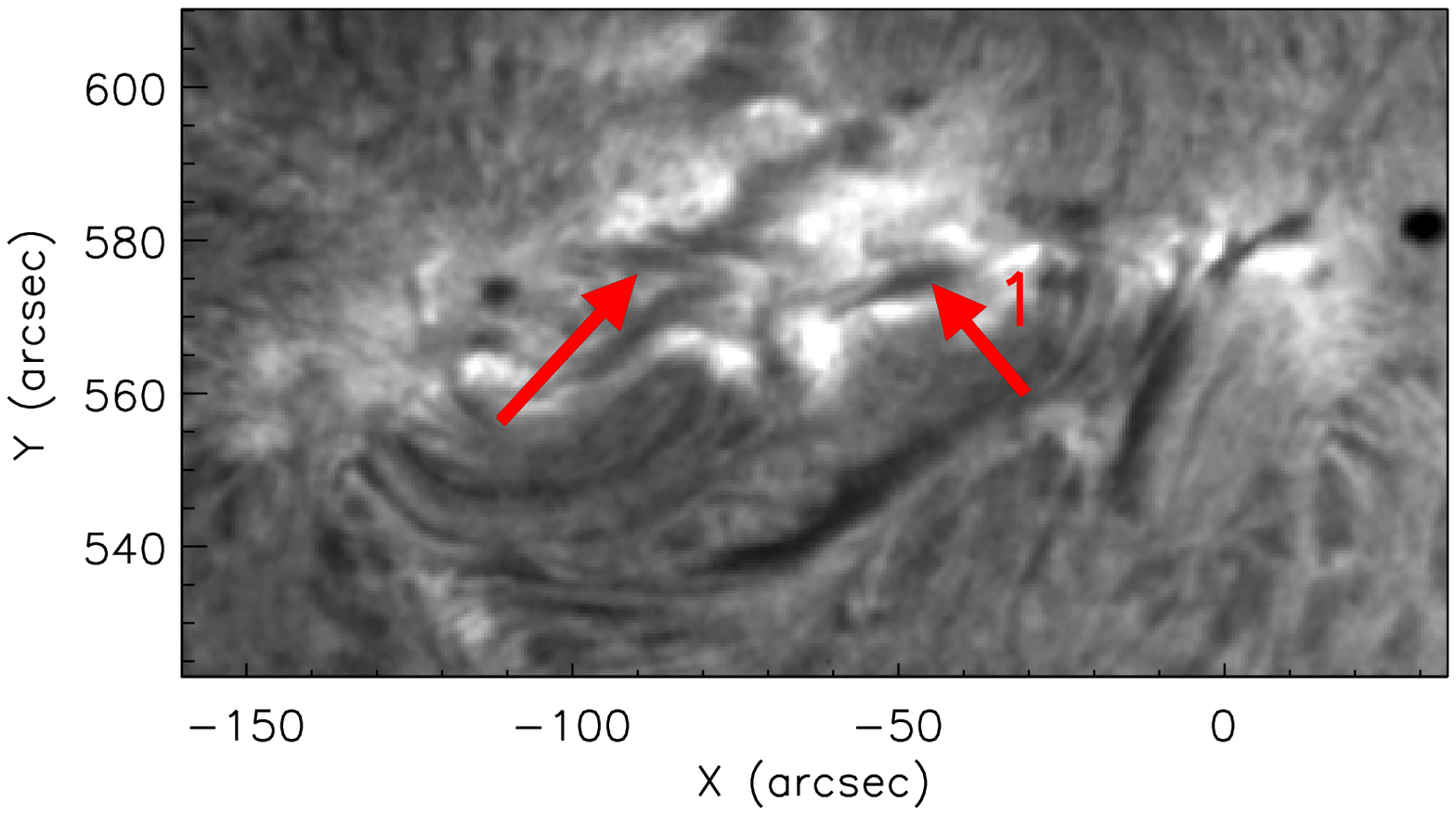}}
  \hspace{0.5cm}
  \subfloat[][28-02-1999]{\includegraphics[width=6.1cm]{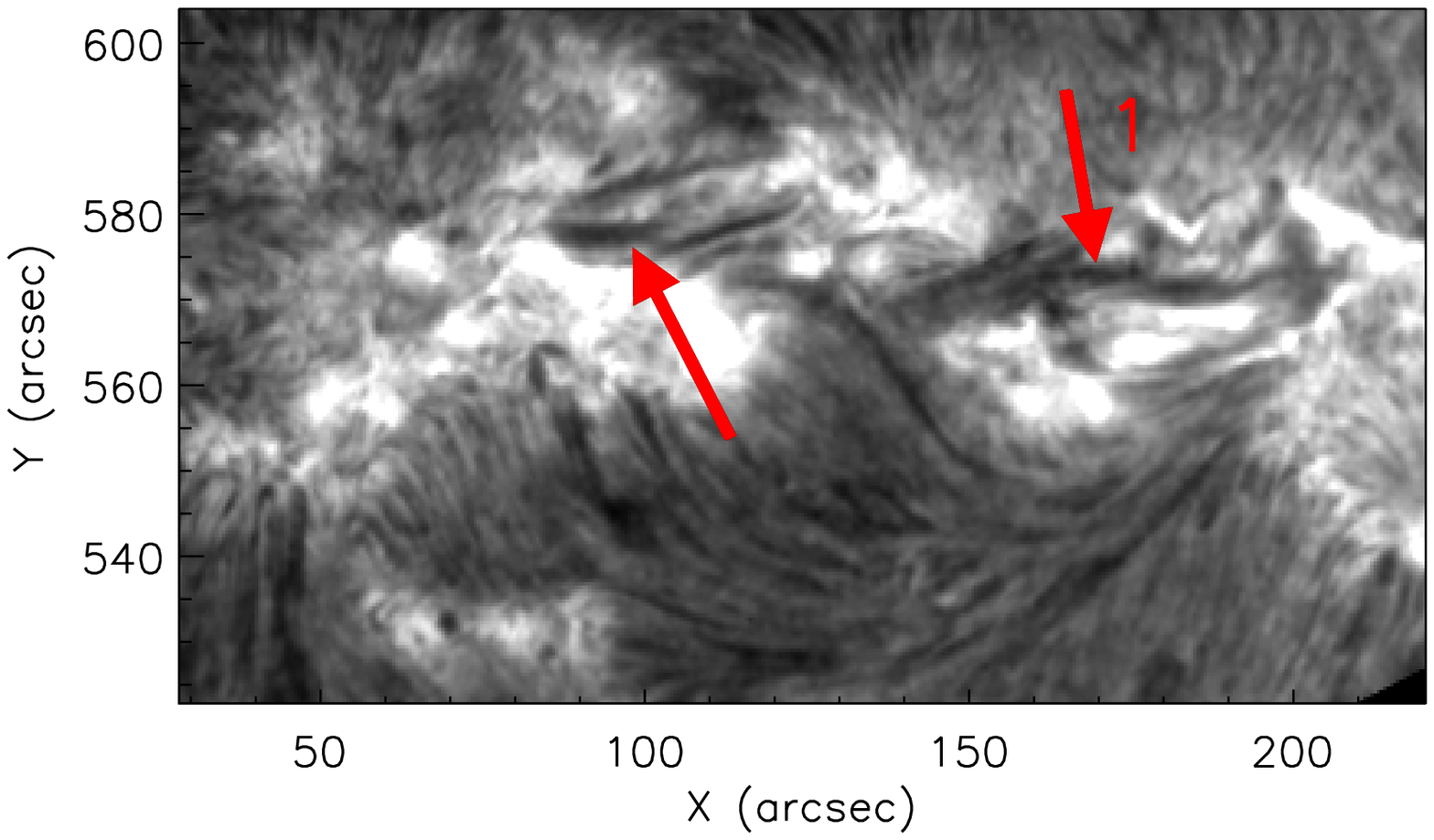}}
    \caption{H$\alpha$ images of NOAA 8471 acquired at BBSO with a pixel size of 1.158 arcsec/pixel. (a): 27-02-1999 at 18:59:24 UT. (b): 28-02-1999 at 17:51:18 UT (after the flare). The field of view is about 148 $\times$ 64 Mm$^2$.}
      \label{evoluc_filament}
\end{figure*}

Figure \ref{evoluc_filament} shows a pair of \ha\ images. The bright facular pattern and several filaments are visible. On 27 February, at 18:59:24 UT (Fig. \ref{evoluc_filament}(a)), a main arched filament (indicated with arrow 1) and several smaller ones can be seen. On 28 February, at 17:51:18 UT (Fig. \ref{evoluc_filament}(b)), after the impulsive phase of the flare, the active region is very bright and the filament 1 and other smaller ones are still visible. 

\subsection{Morphology of the event}

\begin{figure}[!th]
\centering
{\includegraphics[width=9.cm]{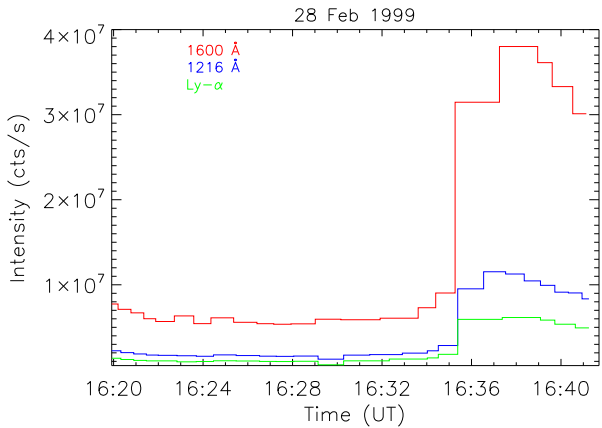}}
    \caption{Temporal evolution of the TRACE 1216~\AA, 1600~\AA\ and corrected ~\lya\ intensities.}
      \label{light_curve}
\end{figure}

Fig. \ref{light_curve} shows the time profiles of the TRACE 1216~\AA, 1600~\AA\ and corrected \lya\ intensity, in counts per second. Note, for the \lya\ intensities we had to time-interpolate the 1600~\AA\ and 1216~\AA\ data to obtain the corrected \lya\ values. Both the 1600~\AA\ intensity and the 1216~\AA\ intensity rise rapidly. From the lightcurves it appears that the 1600~\AA\ intensity peaks later than the 1216~\AA\ intensity; however, the temporal sampling is rather poor and higher-cadence observations are needed to refine this timing. During the decay phase, the intensity at 1216~\AA\ decreases less rapidly than at 1600~\AA. The 1216~\AA\ flare intensity, and therefore also the corrected \lya\ intensity, is substantially smaller than the 1600~\AA\ intensity throughout the event. Both are enhanced above their pre-flare values by a factor of approximately 6. Comparing with the limited previous observations that are available, the 1216~\AA\ intensity increase is consistent with that observed in integrated \lya , in the smaller of the two flares reported by \cite{1980SoPh...67..339C}. The 1600~\AA\ intensity increase is also consistent with that observed in the centre of the C IV 1548~\AA\  line, using the Ultraviolet Spectropolarimeter on SMM \citep{1981ApJ...244L.133W}. C IV 1548~\AA\ is a major line contribution in the TRACE 1600~\AA\ passband. It is therefore not unreasonable to suppose that in the TRACE UV flare data reported in the present work the C IV and Si II lines also contribute substantially to the increases in the broad filter intensities \citep{1999SSRv...87..161D}.

\begin{figure*}
\centering
   \includegraphics[width=18cm]{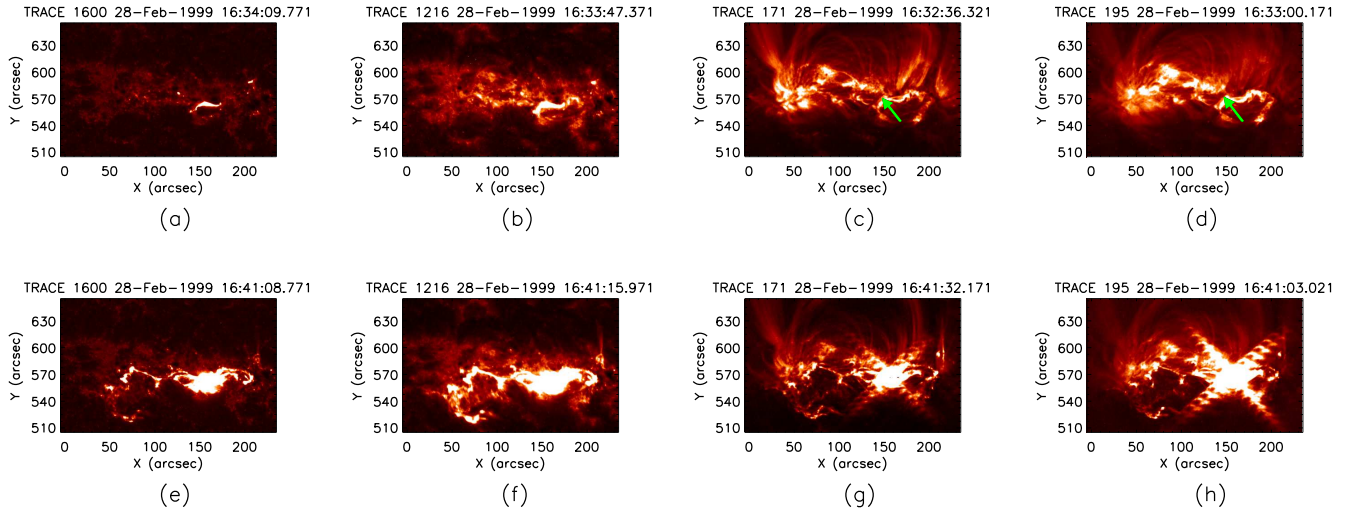}
    \caption{1600~\AA, 1216~\AA, 171~\AA\ and 195~\AA\ TRACE images at the beginning of the impulsive flare phase (a), (b), (c), (d) and at the maximum of the flare (e), (f), (g), (h). The field of view is 184 $\times$ 115 Mm$^{2}$.}
      \label{TRACE_flare2}
\end{figure*}

Figure \ref{TRACE_flare2} shows the signatures in the early stage of the impulsive phase (upper row) and at the peak (bottom row) in the 1600 \AA, 1216 \AA, 171 \AA\ and 195 \AA\ wavelengths, respectively. In the early stage of the impulsive phase (at $\sim$ 16:34 UT), TRACE 1600 ~\AA\ and 1216 ~\AA\ images (Fig. \ref{TRACE_flare2}(a) and (b)) show a sudden brightness increase at location [140:170; 550:560], while the 171~\AA\ and 195~\AA\ images (Figs. \ref{TRACE_flare2}(c) and (d)) show bright areas also in the eastern side of the active region.  

In Fig. \ref{TRACE_flare2} (e), (f), (g) and (h) we can see the increased emission at the peak of the flare (16:41 UT) in different layers of the atmosphere: the TRACE 1600~\AA\ (Fig. \ref{TRACE_flare2}(e)) shows the flare configuration in the chromospheric layer, TRACE 1216~\AA\ in the upper chromosphere (Fig. \ref{TRACE_flare2}(f)), and TRACE 171~\AA\ and 195~\AA\ (Figs. \ref{TRACE_flare2}(g) and (h)) are associated with the transition region and the corona.

At the peak of the impulsive phase the bright emission extends over a greater area [50:200; 520:580] at all wavelengths. We note also the presence of a supergranular cell, the boundaries of which are visible in early 1600~\AA\  and 1216~\AA\  images (\ref{TRACE_flare2}(a) and (b)), spanning approximately [50:100; 520:560]. The south-eastern boundary of this cell becomes enhanced during the flare; this appears to be associated with a flare-related spray (see Figure \ref{evol_plasma_flare2} and Section \ref{sect:disc}).

\subsection{Flare evolution in the \ha\ line center}

\begin{figure*}
\centering
\hbox{
 \hspace{1.cm}
\subfloat[][]{\includegraphics[width=5.0cm]{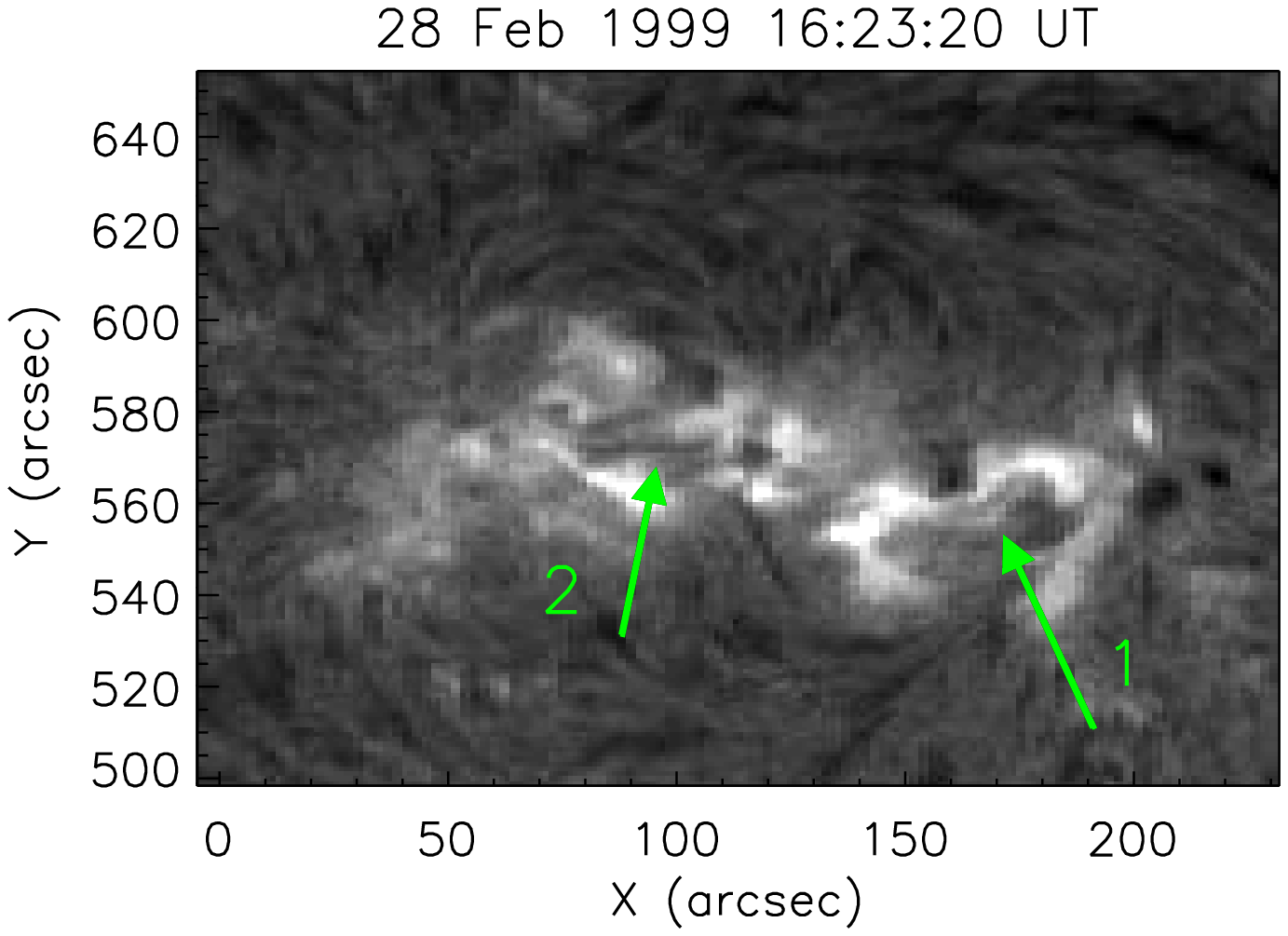}}
 \hspace{0.3cm}
\subfloat[][]{\includegraphics[width=5.0cm]{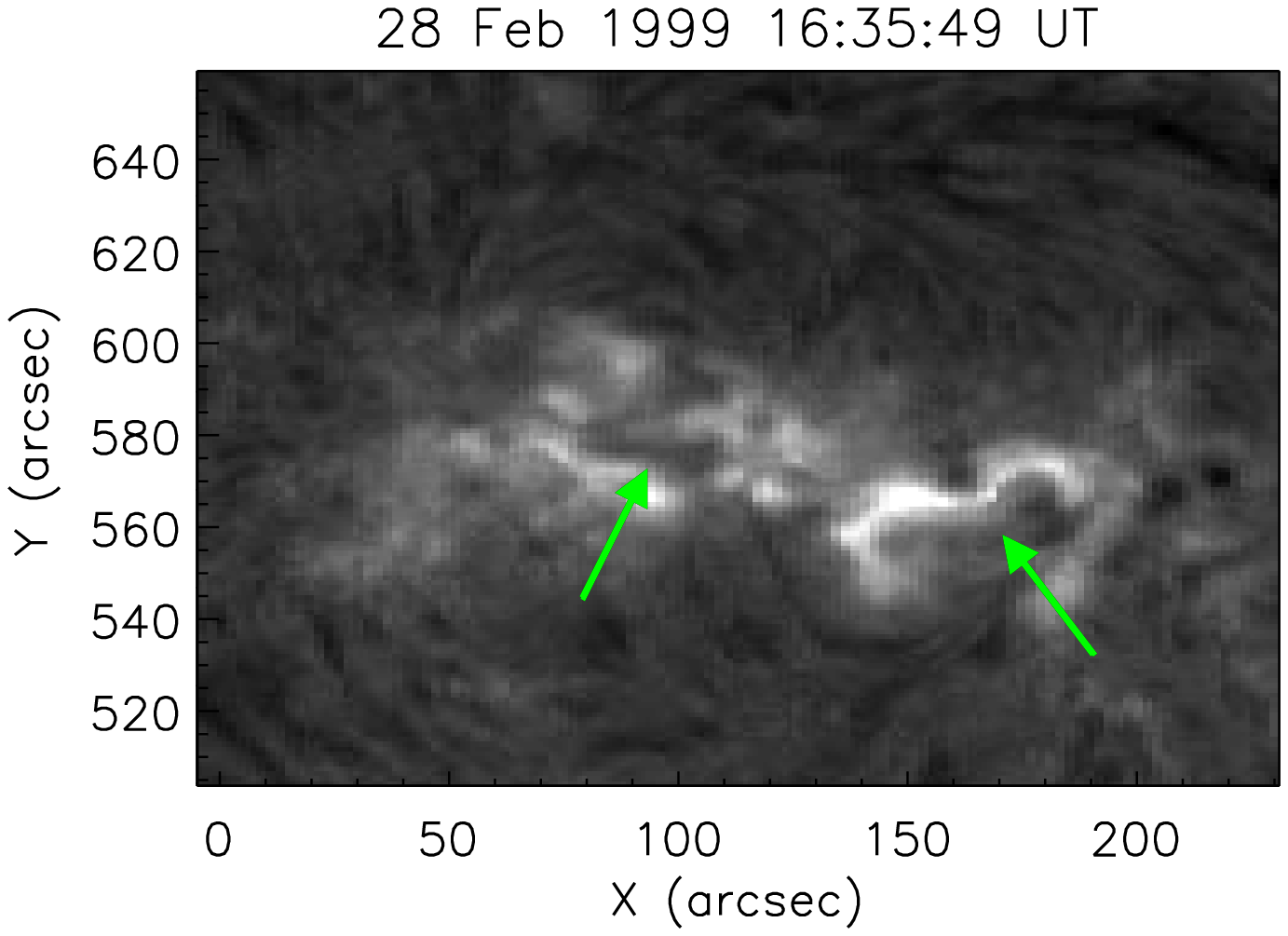}}
 \hspace{0.3cm}
\subfloat[][]{\includegraphics[width=5.0cm]{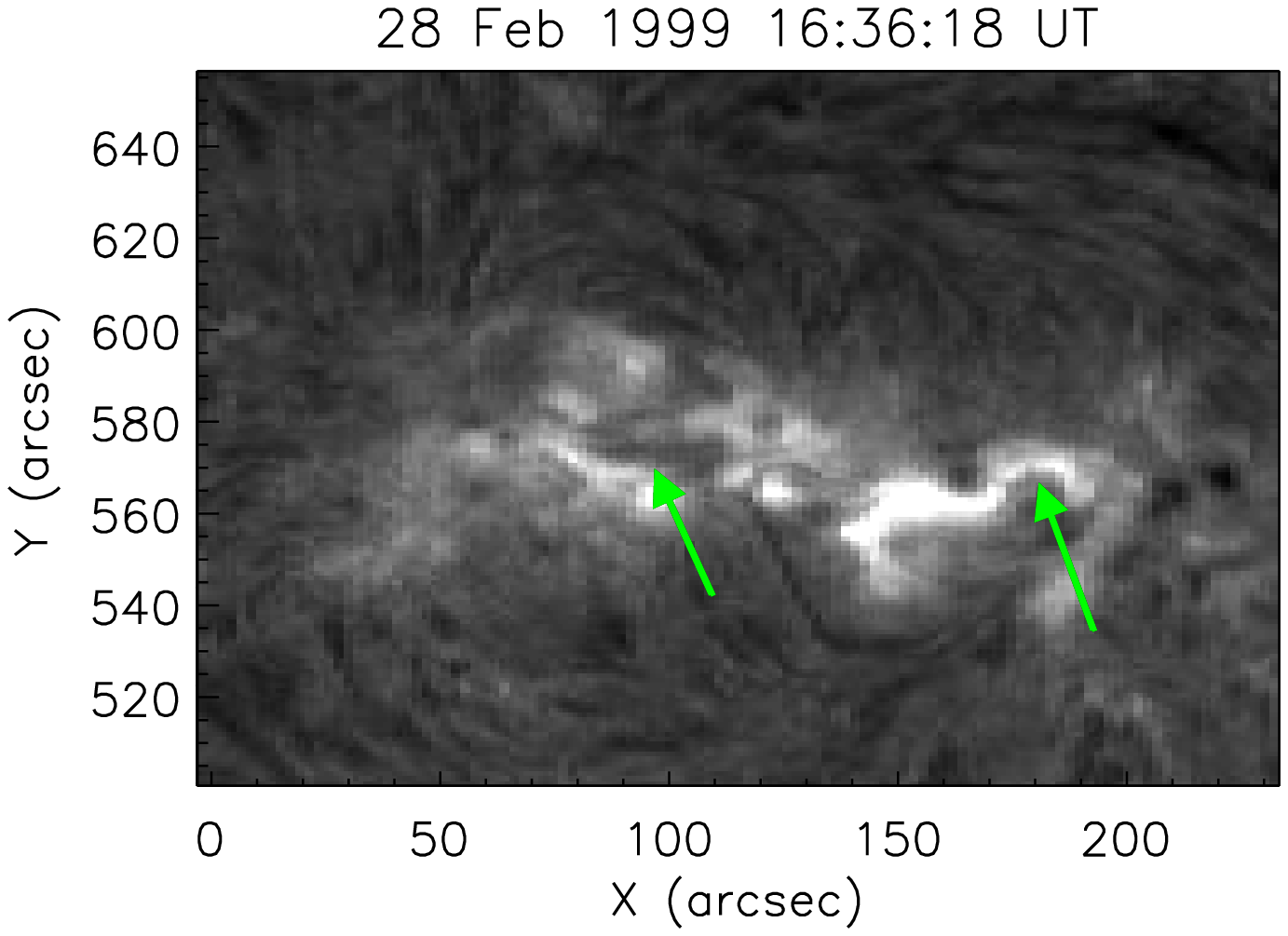}}
}\hbox{
\hspace{1.cm}\vspace{0.3cm}
\subfloat[][]{\includegraphics[width=5.0cm]{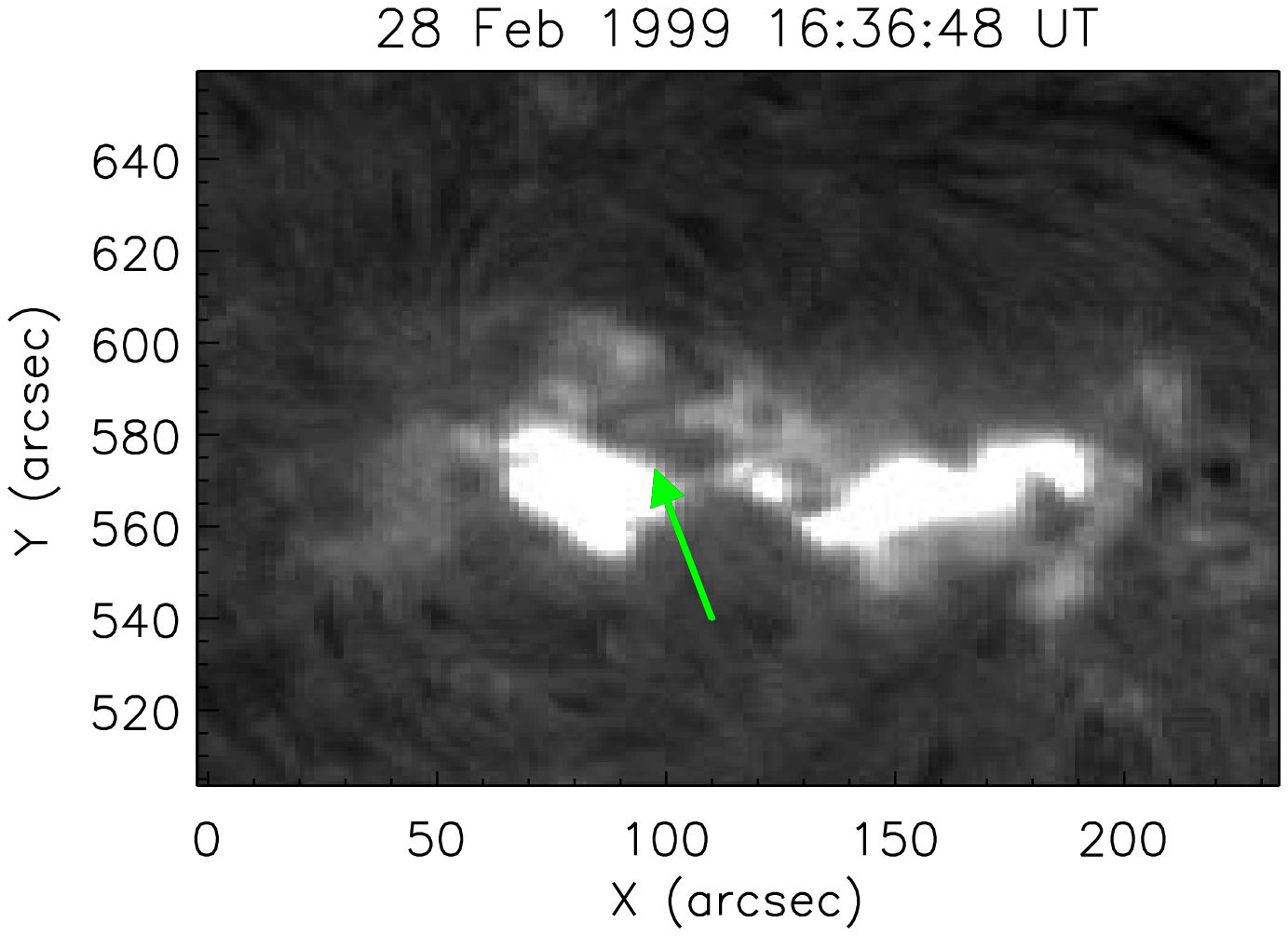}}
 \hspace{0.3cm}
\subfloat[][]{\includegraphics[width=5.0cm]{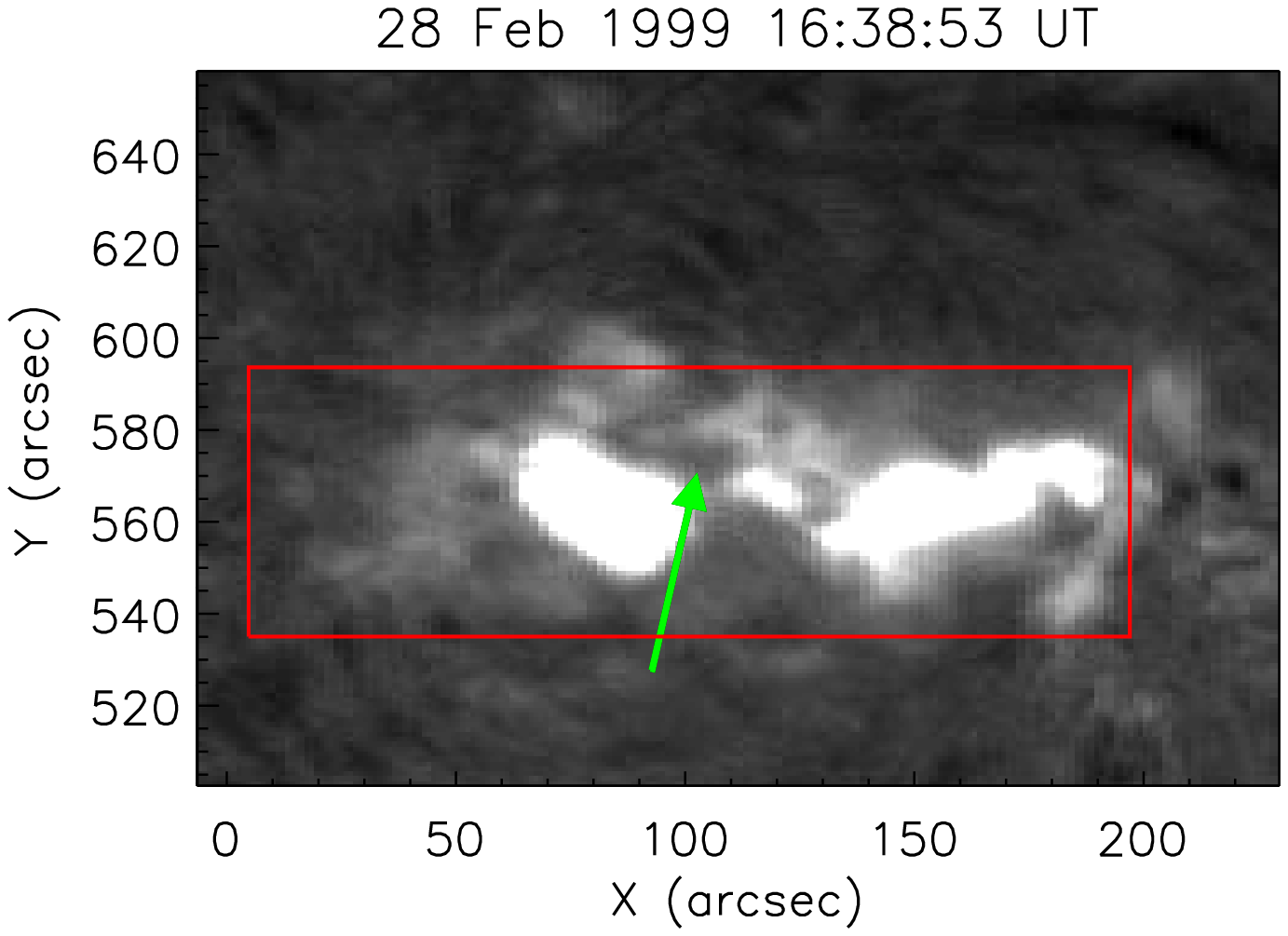}}
 \hspace{0.3cm}
\subfloat[][]{\includegraphics[width=5.0cm]{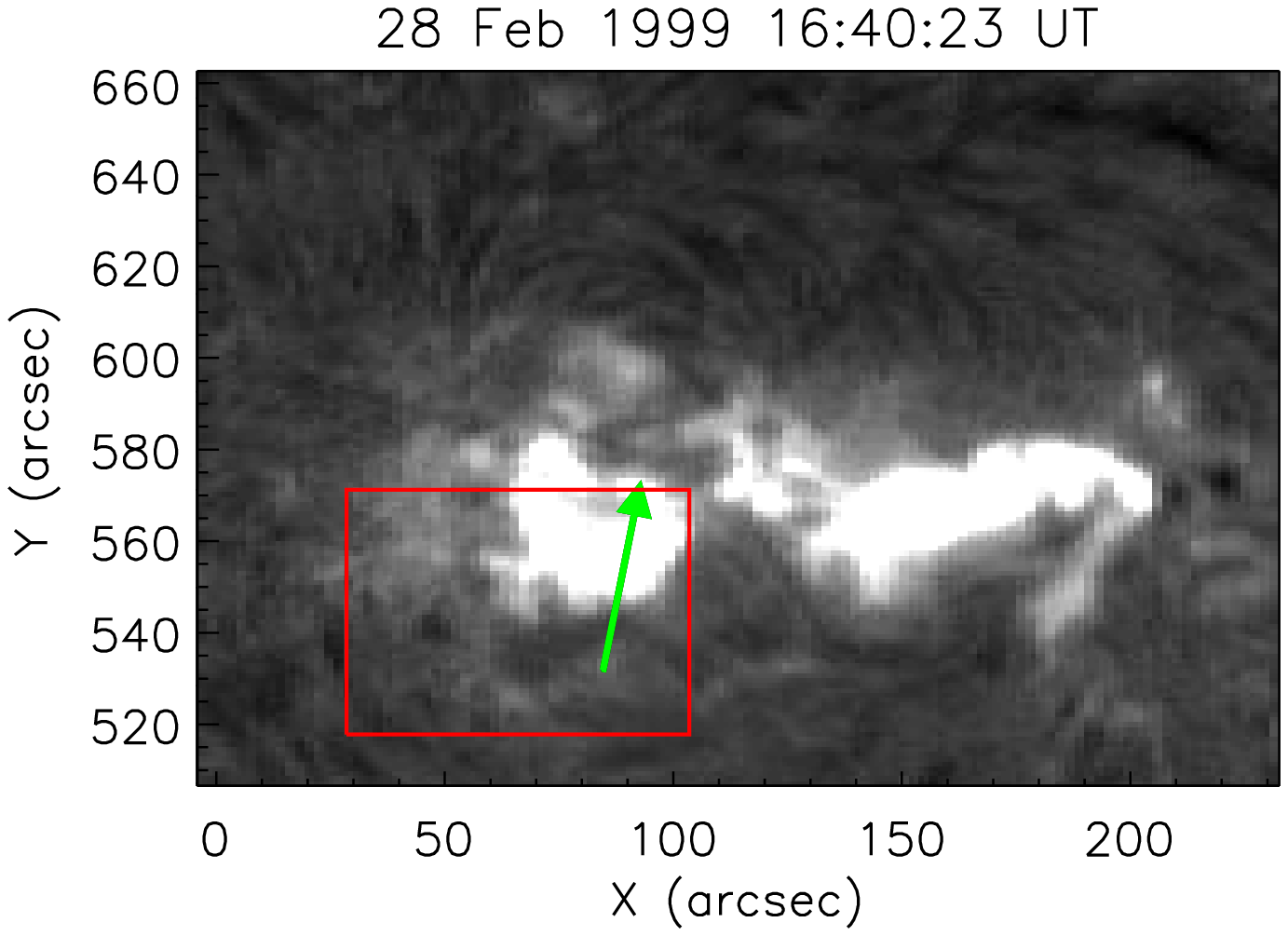}}
}\hbox{
\hspace{1.cm}\vspace{0.3cm}
\subfloat[][]{\includegraphics[width=5.0cm]{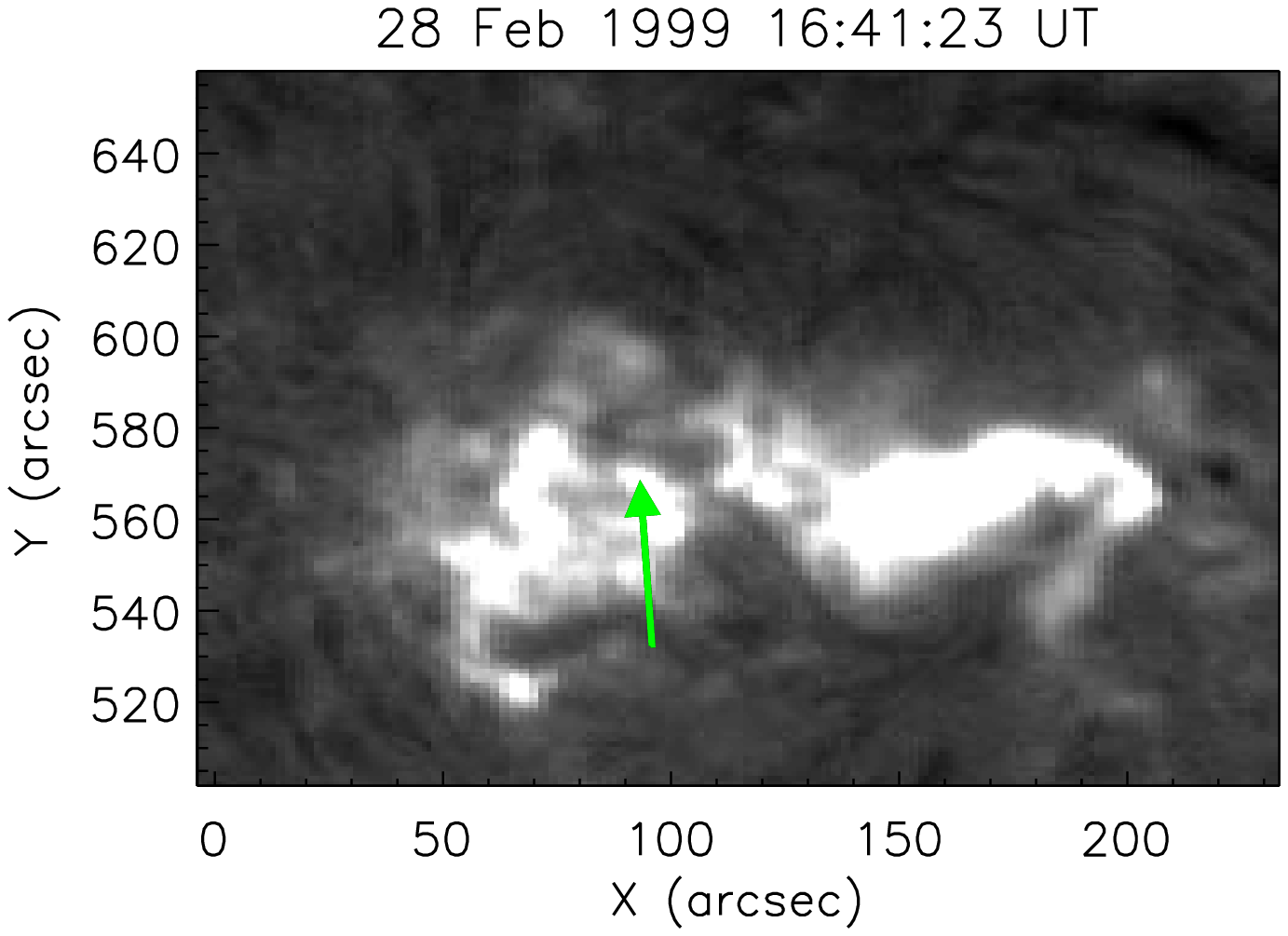}}
 \hspace{0.3cm}
\subfloat[][]{\includegraphics[width=5.0cm]{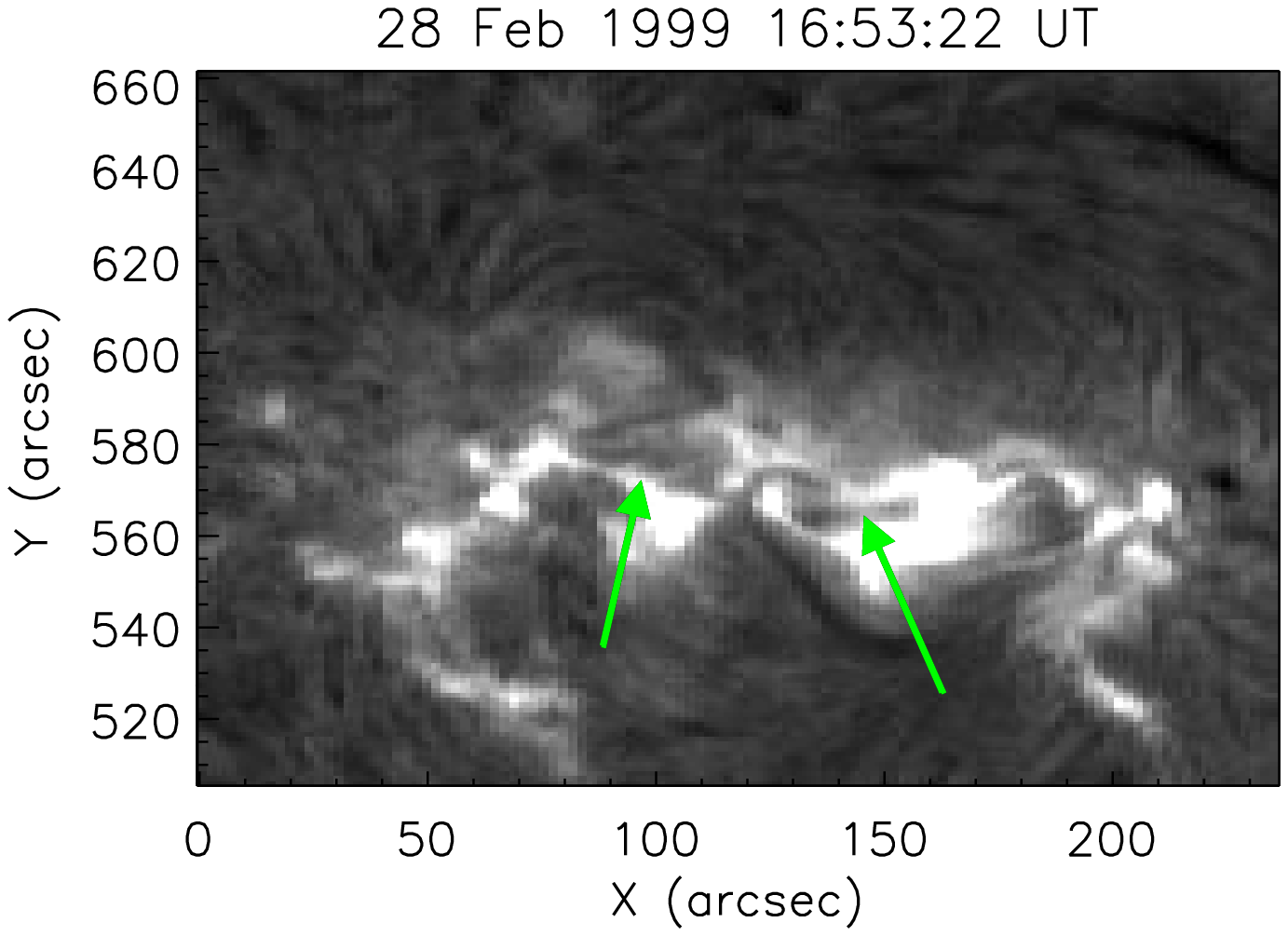}}
 \hspace{0.3cm}
\subfloat[][]{\includegraphics[width=5.0cm]{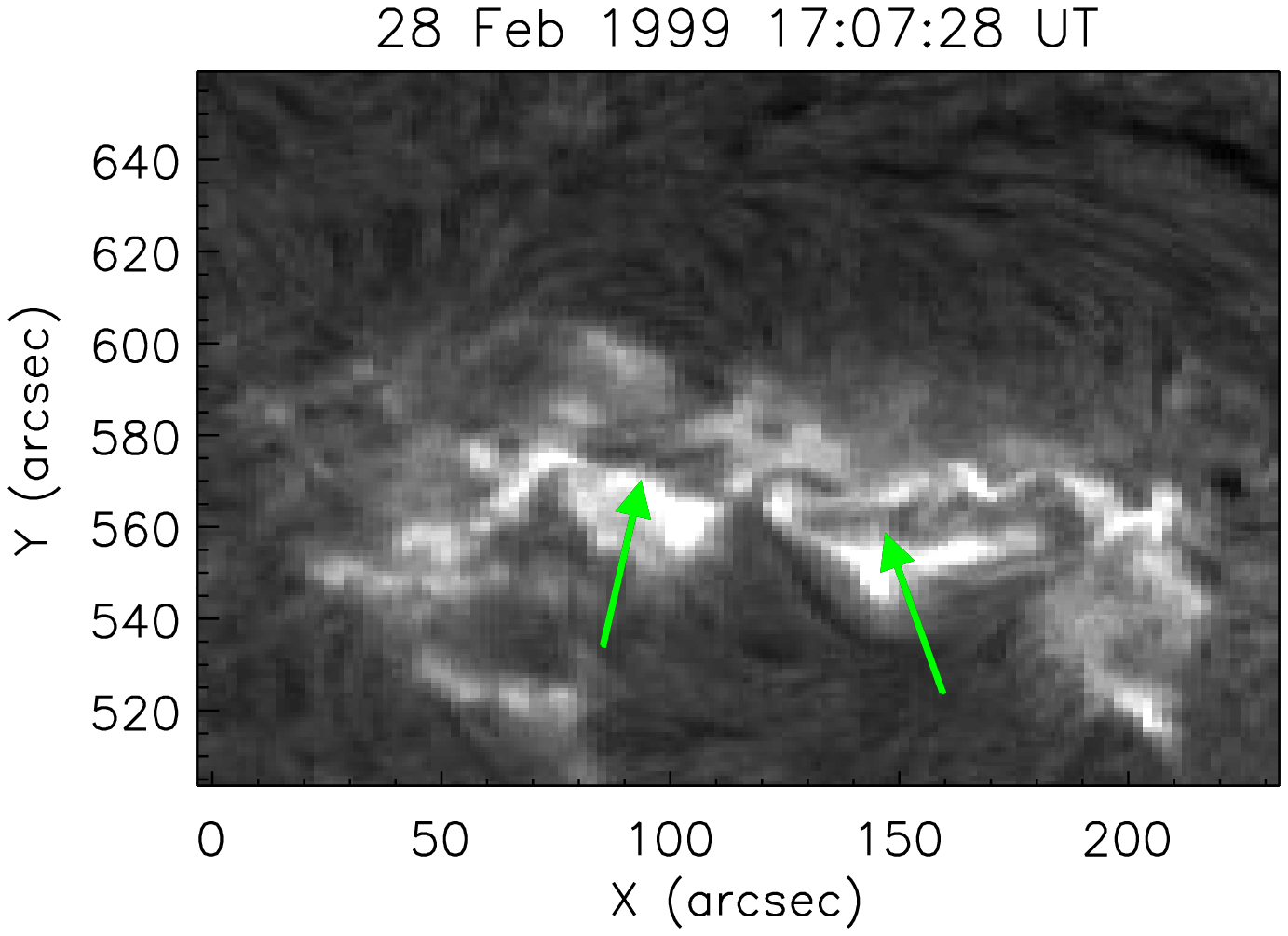}}
}
    \caption[\ha flare evolution]{Flare evolution from 10 minutes before the beginning of the impulsive phase to 40 minutes after the flare in \ha, using a selected area of full disk BBSO images; the green arrows indicate the position of the two main filaments associated to the flare. The boxes in (e) and (f) indicate areas of increased brightness during the flare impulsive phase (see text). The field of view of the images is 182 $\times$ 120 Mm$^2$.}
      \label{evol_halpha}
\end{figure*}

Fig. \ref{evol_halpha} reports the flare evolution in the chromosphere, in the center of the \ha\ line, using BBSO full disk images. Fig.  \ref{evol_halpha}(a) shows the active region about 8 minutes before the beginning of the impulsive phase. We can see that some filaments are present inside the bright facular region. Comparing with the BBSO magnetogram  in Fig. \ref{config_flare2}(c), these filaments outline the main polarity inversion line. We highlight the presence of the two filaments indicated by arrows in Fig. \ref{evol_halpha} (a), from now called filament 1 and filament 2. These features will have a key role in the flaring process.

In the next image (Fig. \ref{evol_halpha}(b) at 16:35:49 UT) the site close to filament 1 brightens, with a clear wave-like pattern. Later on the bright area increases its size and extends toward east, until it reaches the location of filament 2. At 16:38:53 UT (corresponding to the time of the peak of the flare according to the X-ray flux measured by GOES10) in the \ha\ images, the two filaments are not visible any more, as the brightness increase is now distributed along all the region marked by the \begin{bf}rectangle\end{bf} in Fig. \ref{evol_halpha}(e).

Starting from 16:40:23 UT, an increasing bright area extends, in the region marked by a rectangle in Fig. \ref{evol_halpha}(f), toward south-east and, starting from 16:53:22 UT (Fig. \ref{evol_halpha}(h)) the region of increased brightness at the site of filament 1 assumes more clearly the shape of two ribbons, that are generally observed on both sides of erupting filaments and that outline the footpoints of a magnetic arcade (compare Fig. \ref{evol_halpha}(h) and Fig. \ref{evol_halpha}(i)).

\subsection{Flare evolution at 1216~\AA~}
\begin{figure*}
\centering
  \includegraphics[width=18.cm]{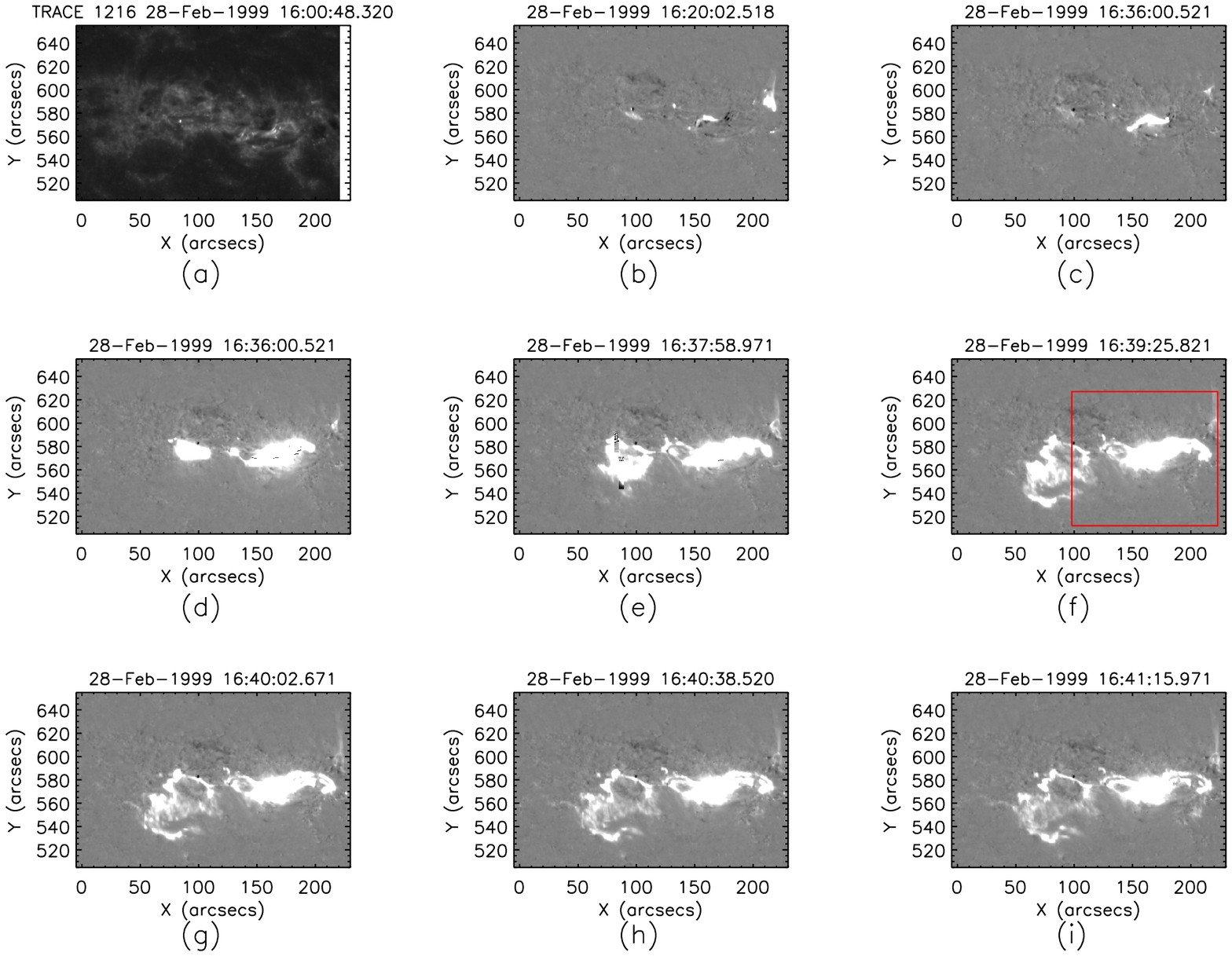}
     \caption{1216~\AA\ base difference images obtained subtracting the image acquired at 16:00:48 UT from the others. The range shown from white to black is $\pm~ 500$ DN/s. The field of view is 180 $\times$ 115 Mm$^2$. The box in Fig. \ref{evol_plasma_flare2}(f) shows the region selected to estimate the \lya\ intensity.}
     \label{evol_plasma_flare2}
\end{figure*}

A series of base difference images from the 1216~\AA\ channel is shown in Fig.~\ref{evol_plasma_flare2} (uncorrected for UV contamination coming from the TRACE 1600~\AA\ emission), where the first image is the base image made at 16:00~UT, $\sim$ 0.5 hour before the beginning of the flare. We have selected only a subset of the images to show the overall evolution. The flat-fielded and normalised TRACE images are first cross-correlated and then the base image is subtracted.
 
At 16:20 UT (Fig. \ref{evol_plasma_flare2}(b)) we can see at the westernmost border of the active region $\sim$ [220; 590] arcsec, a very bright structure resembling a vertical upward jet of plasma. At the same time, at position [140:170; 560:575] arcsec, is a brightness increase having a wave-like shape (compare with the same position in Figs \ref{TRACE_flare2} (a)-(d) and with Fig. \ref{evol_halpha}(b)). In Fig. \ref{evol_plasma_flare2}(c), the bright wave-like feature is still visible. This structure becomes more extended in the following images and in 1 minute and 15 seconds, (See Fig. \ref{evol_plasma_flare2}(c) and Fig. \ref{evol_plasma_flare2}(d)), an increase of brightness appears also in the region [79:106, 567:579], growing in time.  Later the bright area increases along the eastern part of the active region [60:112, 525:590], assuming a very diffuse pattern as time proceeds (figs. \ref{evol_plasma_flare2}(e) and \ref{evol_plasma_flare2}(f).)

This diffuse emission has the morphology of a spray, as also seen in the flare observations presented in \citet{2009A&A...507.1005R}, though in that event the spray material appeared to escape the active region. Due to its proximity to the filament and to its evolution with respect to filament 1 activation, we conclude that the spray emission is associated with the filament activation. In this event, the size of the spray appears to be limited, possibly by the magnetic environment into which it is ejected (see Section \ref{sect:magnetic_field}).

The H${\alpha}$ and TRACE images showed the appearance of a jet-like structure at the western boundary of the active region, at position $\sim$ [220, 590] arcsec (see, e.g. the image at 1216~\AA\ in Fig. \ref{evol_plasma_flare2} (b)). This feature appeared with some intermittency in the intensity before and during the flare. It will be described further in Section \ref{sect:jet_evol}.

\subsection{Estimation of the \lya~intensity at the flare footpoints}
We corrected the 1216~\AA\ TRACE image for the UV contamination. We selected a square around the flare region (from 98 to 223 arcsec in x and from 512 to 627 arcsec in y, shown by a box in Fig. \ref{evol_plasma_flare2}(f)) in a 1216~\AA\ TRACE image at the impulsive phase (16:38:45 UT), for a threshold range between 1200 and 4090 DN, to avoid the saturated pixels. After calibration we measured the 1216~\AA\ intensity in DN units at the flare footpoints (1.3 $\times$ 10$^7$ DN). Taking into account the response of the TRACE filter, the image exposure time, and assuming isotropic emission and that all the energy comes from \lya\ photons, the total power at the flare footpoints in the TRACE 1216~\AA\ is 1.4 $\times$ 10$^{26}$ erg s$^{-1}$ at the impulsive phase. Dividing this value by the number of pixels in the region selected, the power is of 2.7 $\times$ 10$^{22}$ erg s$^{-1}$ pixel$^{-1}$. This value is an upper limit, because there is some contribution from the other UV emission in the 1216~\AA\ filter.
For TRACE 1600~\AA\ image at 16:38:37 UT the power calculated at the footpoint region is 9.0 $\times$ 10$^{22}$ erg s$^{-1}$ pixel$^{-1}$, resulting in a total power is 4.2 $\times$ 10$^{26}$ erg s$^{-1}$.

As explained in \citet{2009A&A...507.1005R}, it is possible to estimate the `pure' intensity of the \lya\ emission line, assuming that the pure \lya\ intensity can be obtained by a linear combination of TRACE 1216~\AA\ and 1600~\AA\ channels \citep{1999SoPh..190..351H}:

\begin{equation}
\centering
I_{Ly\alpha}=A\times I_{1216}+B\times I_{1600},
\label{eq1}
\end{equation}

where $I_{Ly\alpha}$ is the intensity of the corrected \lya\ emission line, $I_{1216}$ and $I_{1600}$ are the intensities as observed with TRACE 1216~\AA\ and 1600~\AA\ channels, respectively, and $A$ and $B$ parameters are obtained by calibrating TRACE data with spectroscopic data from SUMER ($A=0.97$ and $B=-0.14$) \citep{2006A&A...456..747K}. However, the $A$ and $B$ coefficients are derived from quiet Sun observations, and it is not clear that they can be applied to flare observations, in which the ratio of contributions in each of the TRACE channels may be different from their quiet Sun values, depending on the line and continuum excitations. The SOLSTICE flare observations of \citet{1996ApJ...468..418B} showed a proportionally much larger increase in C IV emission than in \lya. The continuum in both regions increased by about a factor 2. Thus the resulting \lya\ intensity may be an overestimate.

Using Eq. \ref{eq1}, the intensity at the flare footpoints in the \lya\ image is corrected for UV leading to a total power in \lya\ at the flare footpoints of 8.1 $\times$ 10$^{25}$ erg s$^{-1}$, a factor 0.6 smaller than that measured in the 1216~\AA\ image. This is equivalent to 1.3 $\times$ 10$^{22}$ erg s$^{-1}$ pixel$^{-1}$. 

The total \lya\ power is more than a factor three greater than was estimated by \citet{2009A&A...507.1005R} for an M1.4 class flare (2.4 $\times$ 10$^{25}$ erg s$^{-1}$, at the beginning of the impulsive phase), as might be expected for a flare that is more energetic in GOES soft X-rays than the M1.4 class flare previously studied (see Table \ref{table_lya_intensity} for a comparison). However, the relationship between GOES class and UV radiation will clearly not be straightforward.  The chromospheric UV radiation emitted will depend on the energy delivered to the chromosphere by flare electrons (and ions), and the GOES class is not necessarily strongly correlated with this \citep{2009A&A...500..901F}.  The characteristics of the UV spectrum generated will also depend on the manner in which this energy is deposited, determined by the details of the electron spectrum and its time evolution, and chromospheric structure. These all influence the resulting temperature, density, ionisation and atomic excitation levels as a function of position. Understanding all this requires detailed calculations, for example radiation hydrodynamics simulations \citep{2005ApJ...630..573A}. It is evidently a fruitful research direction since the lower atmosphere flare UV radiation contains a large amount of diagnostic information.

\begin{table}
\centering
\caption{Intensity calculated in the flare footpoints for both flares.}
\begin{tabular}{c c c}
\hline
 & M6.6 Flare & M1.4 Flare \\
\hline\hline
$I_{1216}$ (erg s$^{-1}$) & $1.4 \times 10^{26}$ & $1.4 \times 10^{26}$\\
$I_{1600}$ (erg s$^{-1}$) & $4.2 \times 10^{26}$ & $7.9 \times 10^{26}$\\
$I_{L_{\alpha}}$ (erg s$^{-1}$) & $8.1 \times 10^{25}$ & $2.4 \times 10^{25}$\\
\hline
\multicolumn{3}{l}{Note: Intensity calculated at the beginning}\\
\multicolumn{3}{l}{of the impulsive phase.}\\
\multicolumn{3}{l}{Note2: We recalculated the values of the}\\
\multicolumn{3}{l}{M1.4 flare \citep{2009A&A...507.1005R}}\\
\multicolumn{3}{l}{taking the same region size and threshold}\\
\multicolumn{3}{l}{range as the M6.6 flare.}\\
\end{tabular}
\label{table_lya_intensity}
\end{table}

\subsection{The jet evolution}\label{sect:jet_evol}
\begin{figure}
\centering
   \includegraphics[width=10.cm]{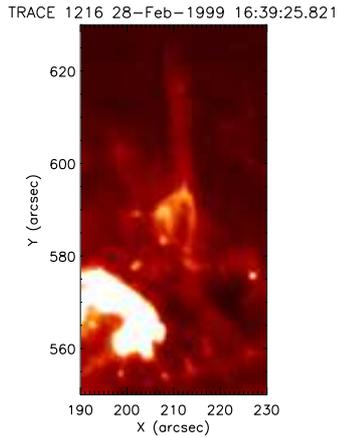}
     \caption{Zoomed image showing the jet at 16:39 UT in the TRACE 1216~\AA~ channel. The field of view of the image is about 31 $\times$ 61 Mm$^2$.}
     \label{evolution_jet_trace_bbso}
\end{figure}

The analysis of the TRACE images showed the appearance of a jet-like structure at the western boundary of the active region, at position $\sim$ [210, 590] arcsec (see, e.g. the image at 1216~\AA~ in Fig. \ref{evol_plasma_flare2} (b)). This feature appeared with some intermittency before and during the flare.
The structure resembles a small loop with a length of $\sim$ 15000 km, a cusped shape and a vertical jet at its top, that can reach an approximate height of $\sim$ 22000 km (see Figs. \ref{evol_plasma_flare2} (b), (e) - (i)). Fig. \ref{evolution_jet_trace_bbso}, shows the jet in the TRACE 1216 \AA~ image acquired at 16: 39 UT.

\begin{figure}
\centering
   \includegraphics[width=9.cm]{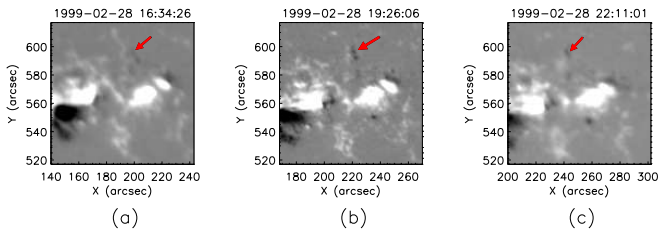}
     \caption{Maps of the magnetic field obtained by the V/I Stokes component acquired at BBSO on 28 February, showing the evolution of the magnetic configuration at the jet site. The arrows indicate the area corresponding to the jet location. The field of view of the images is 78 $\times$ 77 Mm$^2$.}
     \label{evolution_jet_bbso_v}
\end{figure}

In Fig. \ref{evolution_jet_bbso_v}  three BBSO magnetograms show the photospheric magnetic field in the western part of NOAA 8471 at 16:34 UT, 19:26 UT and 22:11 UT. At the location indicated by the arrows, corresponding to the jet-like emission, several knots of positive and negative magnetic field initially approach each other and later decrease in strength and size. It seems plausible that the jet-like structure is due to magnetic reconnection causing a re-arrangement of the magnetic field. The process causes plasma emission at all wavelengths observed, indicating that the plasma at chromosperic and coronal temperatures is involved in the jet.

\section{The Magnetic Configuration}\label{sect:magnetic_field}

\begin{figure*}
\centering
\subfloat[][White Light]{\includegraphics[width=9.cm]{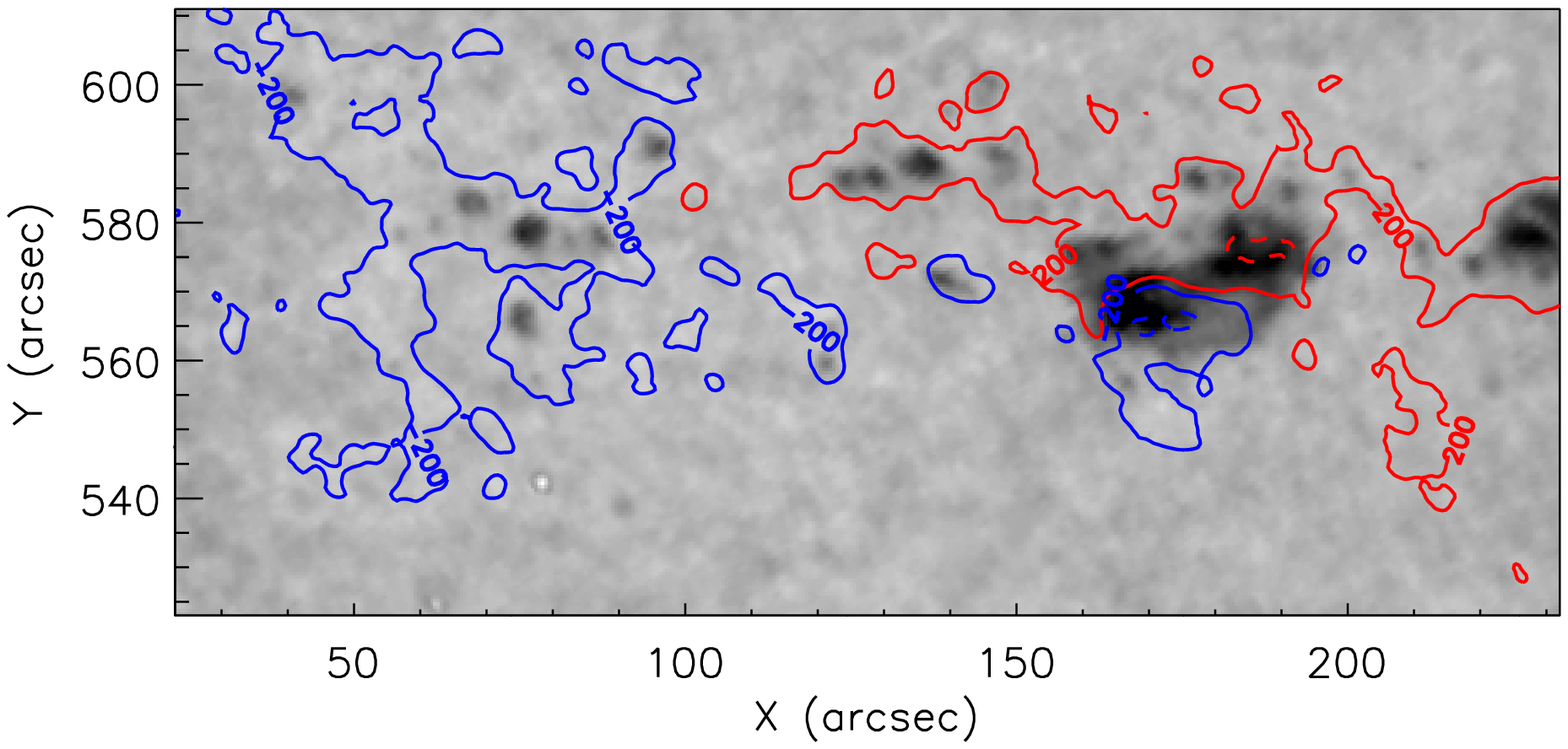}}
\subfloat[][H$\alpha$]{\includegraphics[width=9.cm]{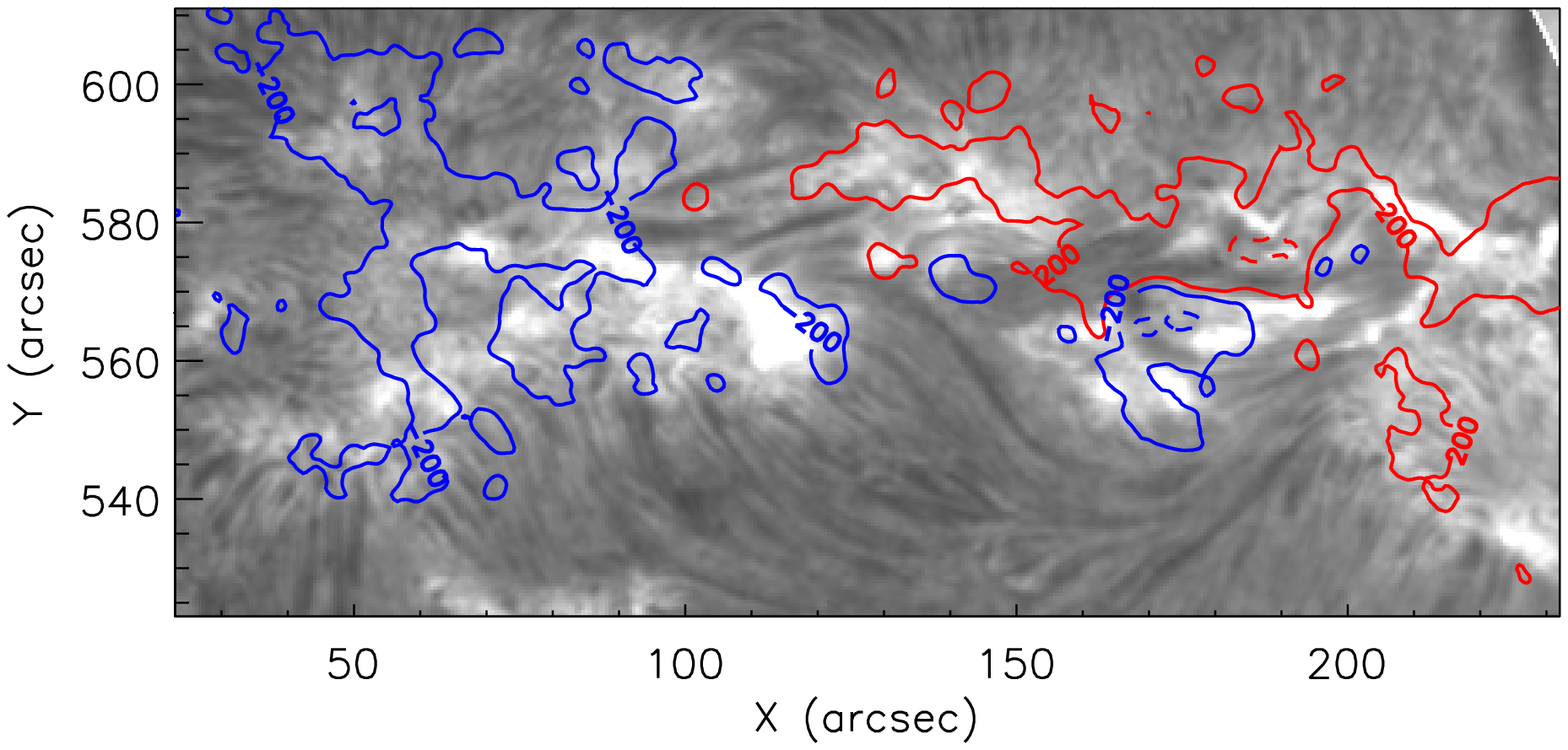}}
     \caption{Isocontours of V Stokes component acquired at BBSO on 28 February at 17:50:31 over (a): the WL image taken at 17:49:52 UT at BBSO; (b) the H$\alpha$ image taken at 17:49:48 UT at BBSO. The red and blue contours indicate the positive and negative magnetic field at $\pm$200 (solid line), $\pm$1000 counts (dashed line). The field of view is of 160 $\times$ 68 Mm$^2$. }
     \label{v_over_wl_ha}
\end{figure*}

In Fig. \ref{v_over_wl_ha}(a), we overplot the BBSO line-of-sight magnetogram over the WL image nearest in time: we can see that the pores in the eastern part of the active region are characterized by negative polarity and that the main sunspot has a $\delta$ configuration, with both positive and negative magnetic polarities inside the same penumbra. Overplotting the BBSO line-of-sight magnetogram over the high resolution \ha\ image nearest in time, we see in Fig. \ref{v_over_wl_ha}(b) that filament 1, the first to become activated, is situated over the photospheric inversion line running along the $\delta$ spot.

Plotting the $\pm$ 200, 1000 G contours of the MDI magnetogram at 16:03:02 UT, before the flare, over the TRACE 171~\AA\ image at 16:30:15 UT, (taking into account the rotation of the sun), we can see in Fig. \ref{mdi_over_171} that the main negative polarity corresponds to the eastern bright region, where higher coronal loops are anchored; the main positive polarity is situated in part over the middle bright region, where the loops are anchored as well and in part in the westernmost boundary.  A bright small loop connects a knot of negative polarity with a diffuse area of positive polarity. The negative knot moved eastward due to shearing motions during the active region evolution.

\begin{figure*}
\centering
  \includegraphics[width=18.cm]{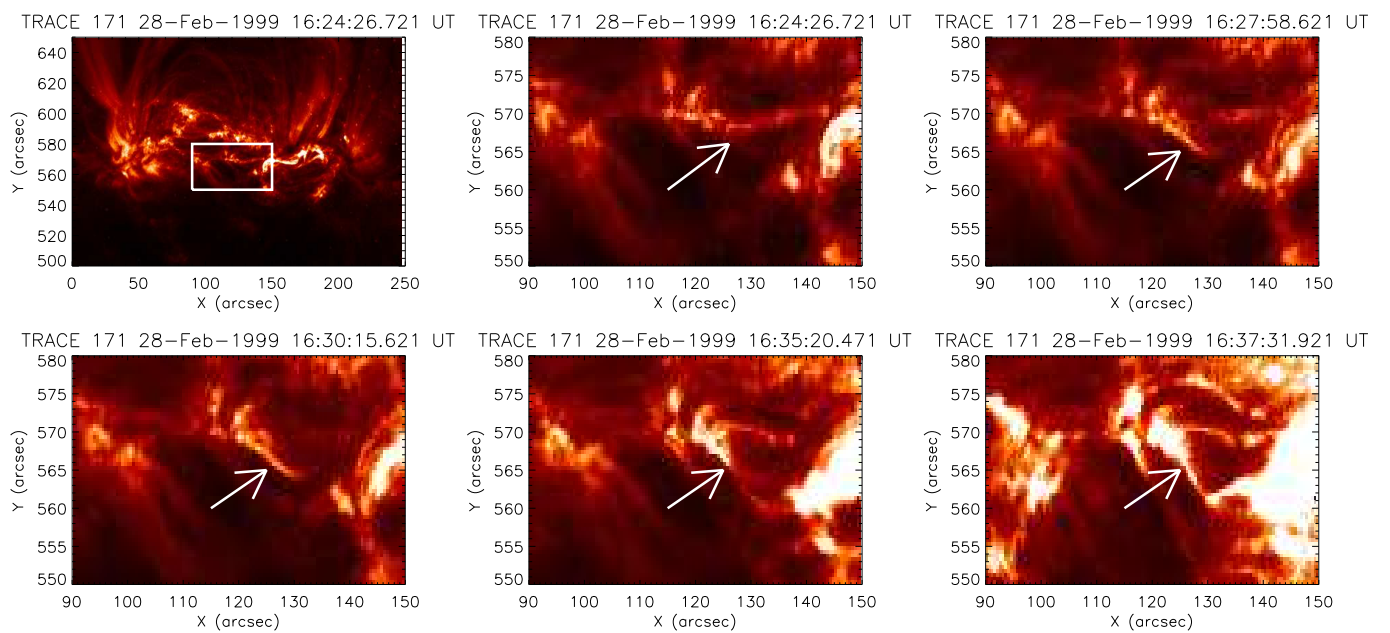}
     \caption{TRACE 171~\AA\ showing the evolution of the flare loop from the beginning of the flare (16:24 UT) till the gradual phase (16:37 UT). The white box in the top left panel corresponds to the area shown in the other panels. The arrows point to the same pixel location in each case, allowing the appearance, growth and motion of the bright loop to be seen.}
     \label{rising_loop}
\end{figure*}

\begin{figure}[!h]
\centering
  \includegraphics[width=10.cm]{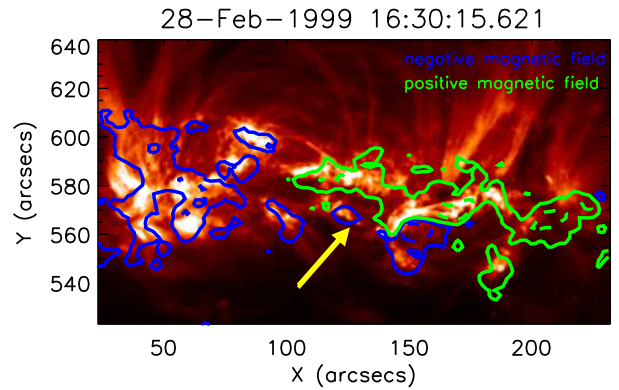}
     \caption{Isocontours of the magnetic field strength deduced from the MDI magnetogram taken at 16:03:02 UT over the TRACE 171~\AA\ image taken at 16:30:15 UT. The green and blue contours indicate the positive and negative magnetic field at $\pm$200 (solid line), $\pm$1000 G (dashed line). The field of view is of 160 $\times$ 90 Mm$^2$. The arrow shows a small bright loop which disappears during the flare (see Fig. \ref{rising_loop}).}
     \label{mdi_over_171}
\end{figure}

Fig. \ref{rising_loop} shows the rising of a flare loop and its association with the onset of brightening in the eastern half of the active region: the top left panel shows the whole active region, for orientation; the other panels are a zoom-in of the central part in the box, and show the small loop that changes. The white box in the top left panel corresponds to the area shown in the other panels. The arrows point to the same pixel location in each case, allowing the appearance, growth and motion of the bright loop to be seen. In the final panel, strong emission starts in the eastern side of the active region. 

\begin{figure*}[!ht]
\centering
\vbox{
 \subfloat[][27 Feb. at 14:27 UT]{\includegraphics[width=6.2cm]{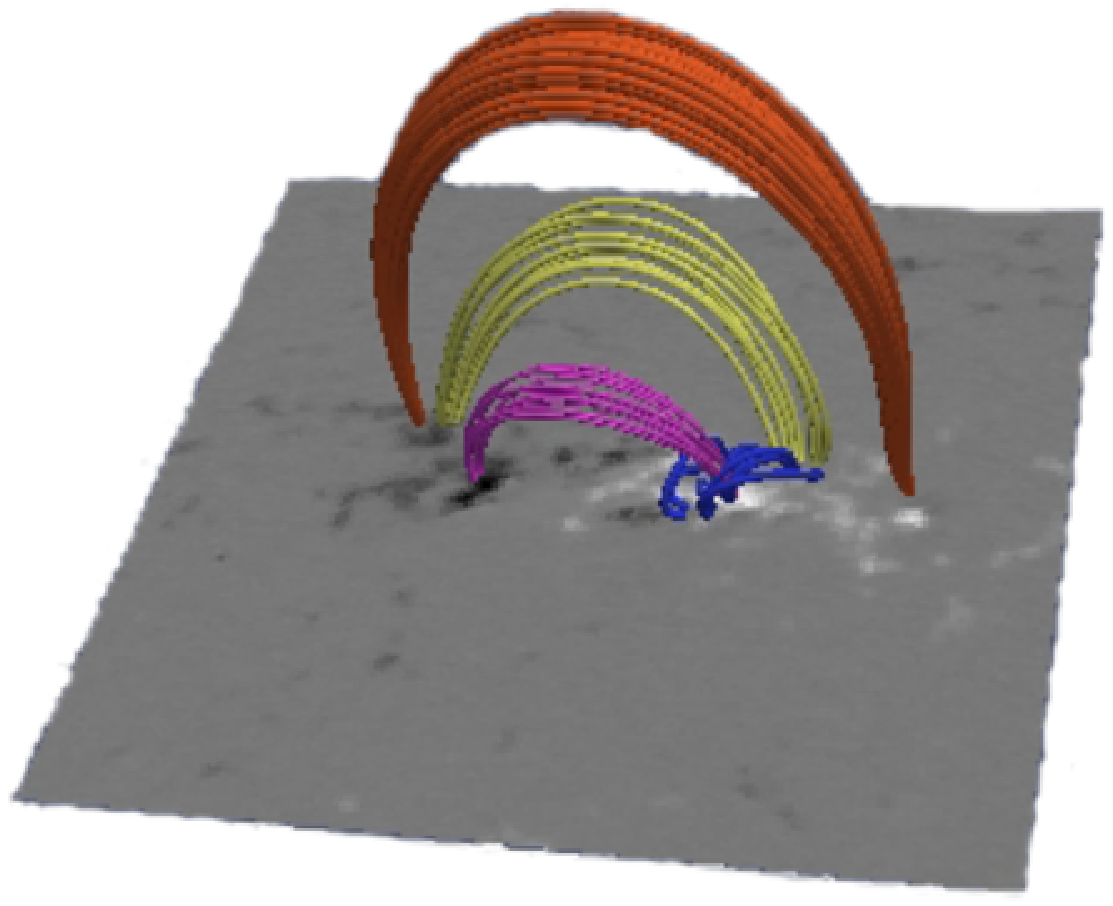}}
 \hspace{1.2cm}
 \subfloat[][28 Feb. at 00:03 UT]{\includegraphics[width=6.2cm]{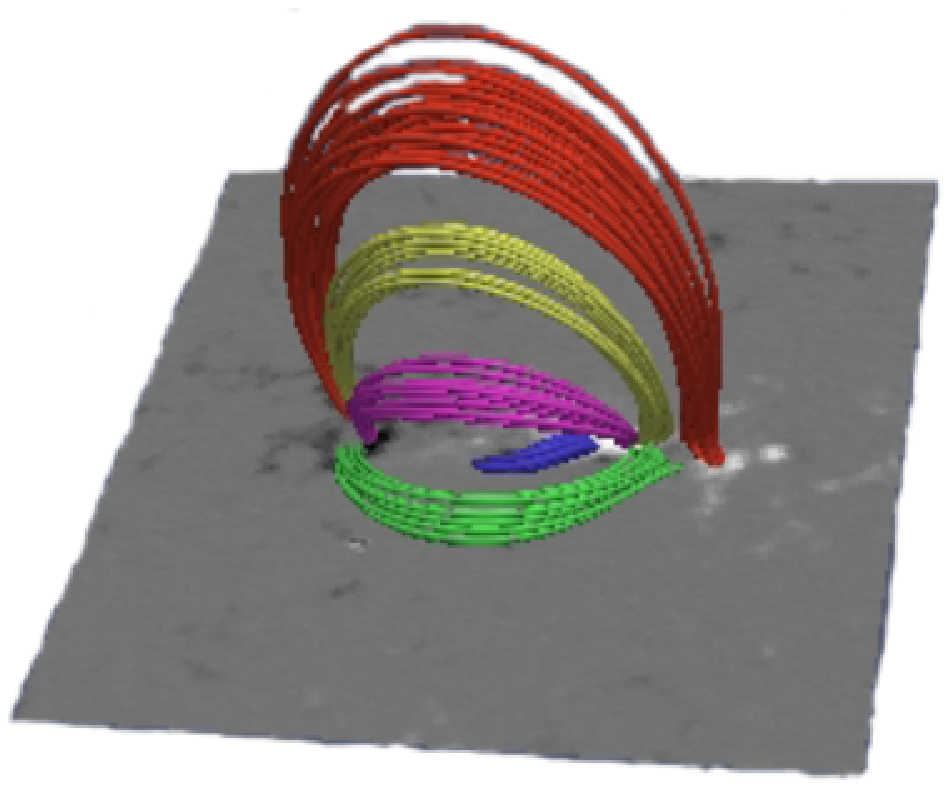}}
}\vbox{ 
 \subfloat[][28 Feb. at 16:03 UT]{\includegraphics[width=6.5cm]{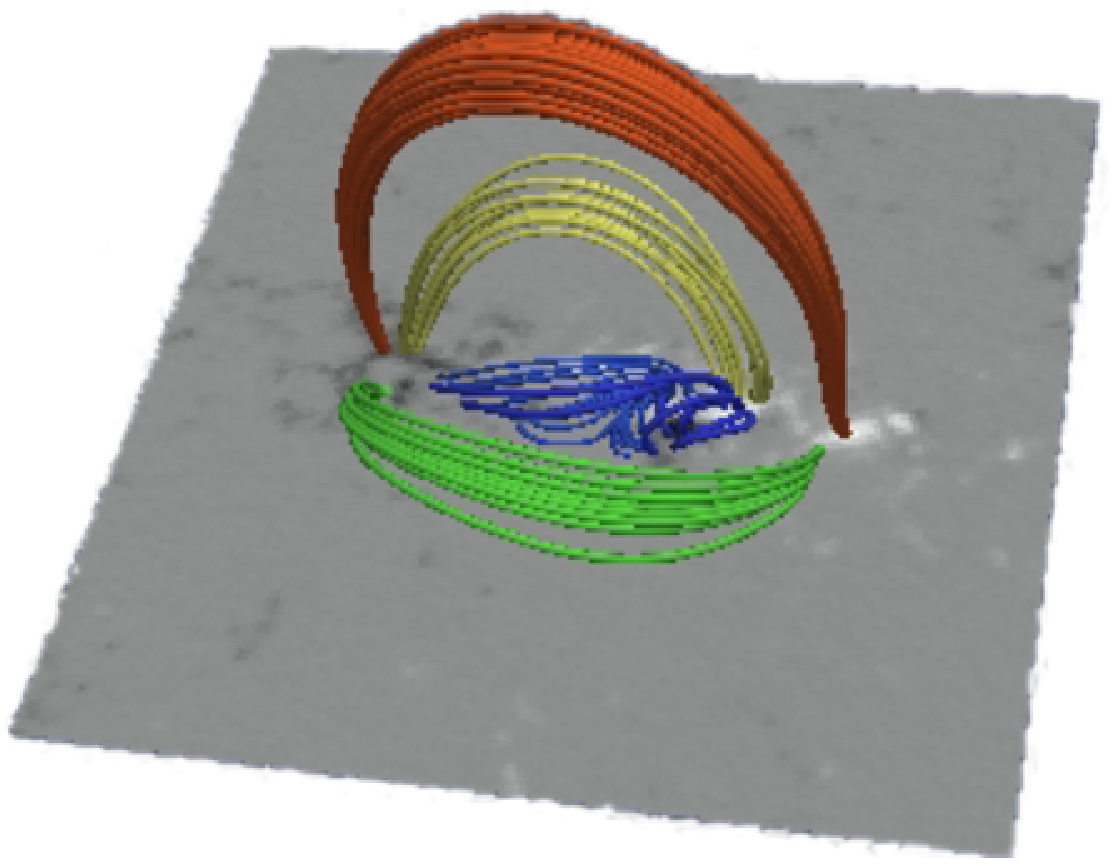}}
 \hspace{1.2cm}
 \subfloat[][28 Feb. at 17:36 UT]{\includegraphics[width=6.5cm]{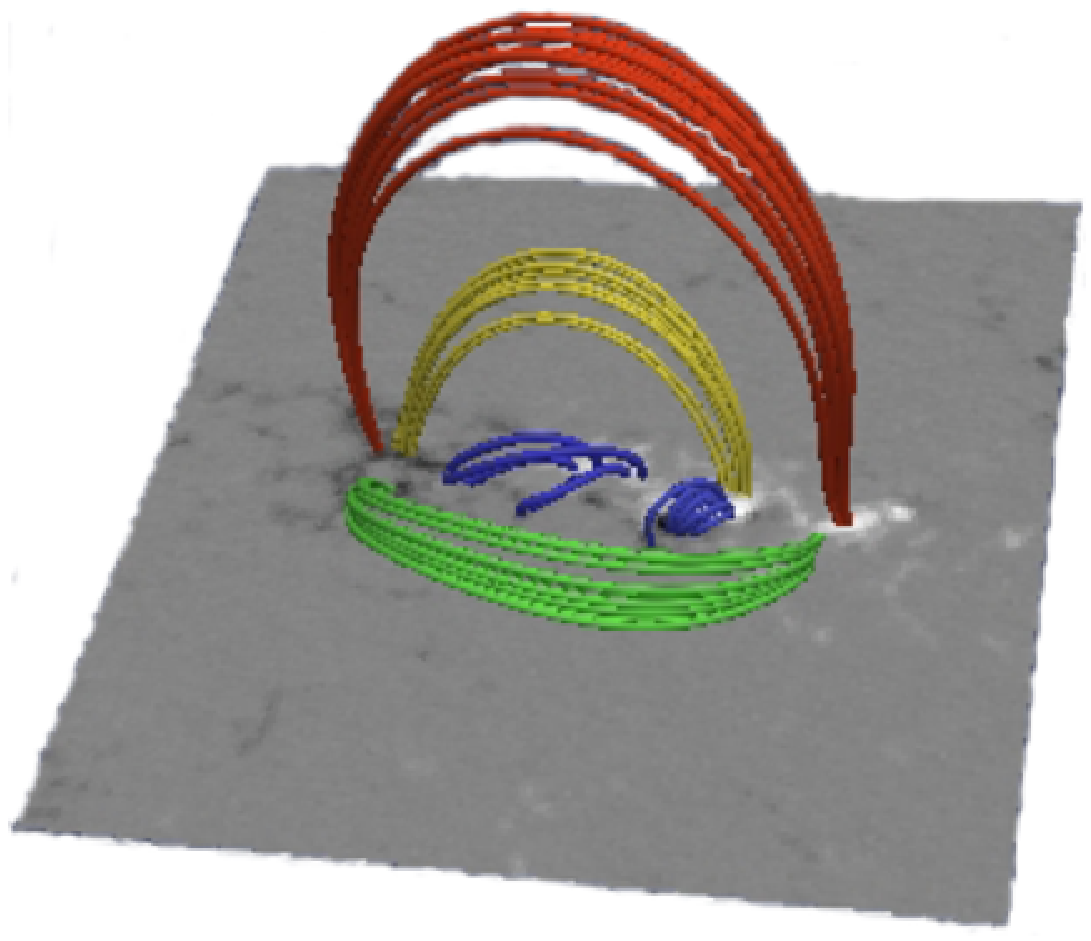}}
 }
     \caption{Schematization of different bundles of field lines inferred from the potential magnetic field extrapolation. The field of view of the background images is about 222 $\times$ 222 Mm$^2$.}
     \label{mag_field3D}
\end{figure*}

We studied the magnetic configuration of the active region using the longitudinal component of the photospheric magnetic field provided by MDI and an extrapolation method introduced by \citet{1981A&A...100..197A}, which allows us to reconstruct the 3D magnetic field above the photosphere (boundary). This method assumes that the magnetic field is force-free not only in the corona but also in the lower levels and that it vanishes at infinity. 

Using a program developed by \citet{2007ApJ...671.1034G}, we computed the potential field (i.e. $\alpha$=0) for the magnetograms taken at 14:27 UT (Fig. \ref{mag_field3D}(a)) on 27 February, at 00:03 UT (Fig. \ref{mag_field3D}(b)), 16:03 UT (Fig. \ref{mag_field3D}(c)) and 17:36 UT (Fig. \ref{mag_field3D}(d)) on 28 February. Even if the potential field approximation is not the most realistic one, it can be helpful to crudely assess the pre- and post-flare magnetic connectivity.

In the extrapolation made using the magnetogram acquired about one day before the event (Fig. \ref{mag_field3D}(a)) we identify four different systems of field lines: a higher arcade (red lines) connecting the easternmost (negative) and the  westernmost (positive) polarities, a lower arcade (yellow lines) connecting the more diffuse magnetic concentrations of both polarities located in the centre of the active region, another bundle of field lines (magenta lines) connecting the most intense negative polarity concentration with the most intense positive one, and some small loops (blue lines) connecting two emerging negative polarities and the main positive concentration in the south-west part of the active region, where the $\delta$ spot appears in the following hours.

Later these two negative emerging polarities moved eastward, as we can see in Fig. \ref{mag_field3D}(b). The blue field lines that connect the same magnetic concentrations appear more stretched due to shearing motions. We also report a green bundle of field lines that is almost parallel to the solar surface. We note that some of these field lines have the same location and reproduce approximately the shape of the long southern loops observed at 171~\AA\ (see Fig. \ref{evol_mdi_trace171} (e)).

When the $\delta$ spot appears, the extrapolations indicate a rearrangement of the magnetic configuration in its south-west part.  A few minutes before the flare (Fig. \ref{mag_field3D}(c)) a new system of field lines corresponding to the $\delta$ configuration appears and the two negative polarities are in part still connected to the main positive polarity and in part are newly connected to the more diffuse positive concentrations in the centre of the active region (in this representation we do not show the magenta field lines, in order to avoid confusion). Finally, after the flare the potential field extrapolation (Fig. \ref{mag_field3D}(d)) shows that the two negative concentrations are connected only to the positive concentrations in the core of the active region and their link to the main positive spot seems to have disappeared.

\begin{figure}[!h]
\centering
 \includegraphics[width=9.cm]{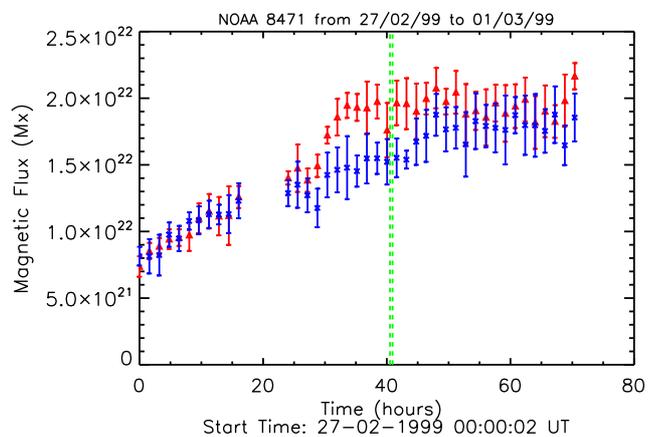}
     \caption{Evolution of the magnetic flux of the active region NOAA 8471 deduced by MDI magnetograms from the day before to the day after the flare. The red and blue symbols indicate the positive and negative magnetic flux respectively and the green vertical line shows the time occurrence of the M6.6 flare.}
     \label{Bflux}
\end{figure}

We also studied the variation of the positive and negative magnetic flux (using MDI/SOHO data, with the same field-of-view as in Fig. \ref{mag_field3D}: 289 $\times$ 289 arcsec around the active region) from the day before the flare (27-Feb-1999) to the day after the flare (01-March-1999). We can see (Fig. \ref{Bflux}) that the magnetic flux increases to a maximum and then remains roughly constant. The green vertical line indicates the time of the M6.6 flare occurrence. The red symbols indicate the positive magnetic flux and the blue ones the negative magnetic flux. Therefore, the flare occurs not at the time of most rapid flux emergence, but once the magnetic structure has fully emerged and stabilised.

\section{Discussion and Conclusions}\label{sect:disc}
We have studied an M6.6 flare using multi-wavelength observations. Image data allowed us to track the evolution of the flare in the context of the active region magnetic configuration provided by BBSO and MDI magnetograms. The TRACE data available at 1600~\AA\  and 1216~\AA\ during the peak of the flare were also used to determine the \lya\ power radiated by the footpoints ($8.1 \times 10^{25}$ erg s$^{-1}$). The observations suggest a sequence of phenomena that can be described in terms of a scenario where a first filament destabilization acts as a trigger for a two-ribbon flare and in about fifteen minutes causes the eruption of a second filament belonging to the same active region. 

More precisely, initially a filament (filament 1) is activated in the westernmost side of the active region. This filament is located along the region separating the opposite polarities of a $\delta$ spot. 

The bright wave-like feature observed both in the chromosphere and the corona reflects the morphology of filament 1, indicating that the plasma forming the filament is heated. This destabilization is probably not due to flux emergence events (see, e.g. the plot in Fig. \ref{Bflux}), but rather it may occur as a result of internal reconnection in the filament or interaction with surrounding field. In particular, we note from Fig. \ref{rising_loop} that a small loop joining the central negative polarity  and the main positive polarity close to the filament, rises up and appears to touch the filament just at the time of brightening. The activity in this small loop indicates that the central negative polarities are also affected by the filament activation, providing a possible link to the eastern part of the active region. One of these negative polarities is located at the end of filament 2, so reconfiguration of the linked field may result in the observed brightening and spray. The shearing motions occurring in the core of the active region probably play a key role in such effects.

The associated brightness increase, having initially  this wave-like shape, broadens along the magnetic inversion line and, while assuming the shape of two ribbons, also propagates toward the eastern part of the active region. Here the second filament (filament 2) is activated approximately three minutes before the flare soft X-ray peak and subsequently develops on a larger area, giving rise to a spray which involves also a quiet Sun area in the south-east part of the active region, propagating along the border of a supergranular cell. It is possible that the expansion of the spray is inhibited by the configuration of low-lying loops to the south of the active region (green lines in Fig. \ref{mag_field3D}). A CME was observed by LASCO/C2 coronograph at 17:54 UT and, according to the height-time measurements, the CME was initiated at around 16:42 UT, that is shortly after the activation of the second filament.

Putting together this sequence of events and the results obtained by the magnetic field extrapolation, we can infer that the second filament destabilization was a consequence of the re-arrangement of the magnetic field line connectivity caused by reconnection occurring in the magnetic field separator over the $\delta$ spot. In particular, the comparison between the magnetic field extrapolation before and after the flare shows a completely different morphology for the lower arcades (blue lines in Fig. \ref{mag_field3D}(a) and (b)) of the active region, therefore confirming the occurrence of the reconnection process. 

Therefore, we can conclude that in this flare both filaments played an \textit{active} and decisive role, because the destabilization of filament 1 caused the reconnection and the re-arrangement of the magnetic field configuration, while the eruption of filament 2 caused the CME.

\begin{acknowledgements}
This work was supported by the European Commission through the SOLAIRE Network (MRTN-CT-2006-035484), STFC rolling grant STFC/F002941/1, by the EC-funded HESPE project (FP7-2010-SPACE-1/263086) and by Leverhulme grant F00-179A.
\end{acknowledgements}

\bibliographystyle{aa}
\bibliography{paper_flare2}

\begin{thebibliography}{28}
\expandafter\ifx\csname natexlab\endcsname\relax\def\natexlab#1{#1}\fi

\bibitem[{{Alissandrakis}(1981)}]{1981A&A...100..197A}
{Alissandrakis}, C.~E. 1981, \aap, 100, 197

\bibitem[{{Allred} {et~al.}(2005){Allred}, {Hawley}, {Abbett}, \&
  {Carlsson}}]{2005ApJ...630..573A}
{Allred}, J.~C., {Hawley}, S.~L., {Abbett}, W.~P., \& {Carlsson}, M. 2005,
  \apj, 630, 573

\bibitem[{{Antiochos} {et~al.}(1999){Antiochos}, {DeVore}, \&
  {Klimchuk}}]{1999ApJ...510..485A}
{Antiochos}, S.~K., {DeVore}, C.~R., \& {Klimchuk}, J.~A. 1999, \apj, 510, 485

\bibitem[{{Aschwanden}(2004)}]{2004psci.book.....A}
{Aschwanden}, M.~J. 2004, {Physics of the Solar Corona. An Introduction}, ed.
  {Aschwanden, M.~J.} (Praxis Publishing Ltd)

\bibitem[{{Brekke} {et~al.}(1996){Brekke}, {Rottman}, {Fontenla}, \&
  {Judge}}]{1996ApJ...468..418B}
{Brekke}, P., {Rottman}, G.~J., {Fontenla}, J., \& {Judge}, P.~G. 1996, \apj,
  468, 418

\bibitem[{{Canfield} \& {van Hoosier}(1980)}]{1980SoPh...67..339C}
{Canfield}, R.~C. \& {van Hoosier}, M.~E. 1980, \solphys, 67, 339

\bibitem[{{Dammasch} {et~al.}(1999){Dammasch}, {Hassler}, {Curdt}, \&
  {Wilhelm}}]{1999SSRv...87..161D}
{Dammasch}, I.~E., {Hassler}, D.~M., {Curdt}, W., \& {Wilhelm}, K. 1999, \ssr,
  87, 161

\bibitem[{{Falewicz} {et~al.}(2009){Falewicz}, {Rudawy}, \&
  {Siarkowski}}]{2009A&A...500..901F}
{Falewicz}, R., {Rudawy}, P., \& {Siarkowski}, M. 2009, \aap, 500, 901

\bibitem[{{Freeland} \& {Handy}(1998)}]{1998SoPh..182..497F}
{Freeland}, S.~L. \& {Handy}, B.~N. 1998, \solphys, 182, 497

\bibitem[{{Georgoulis} \& {LaBonte}(2007)}]{2007ApJ...671.1034G}
{Georgoulis}, M.~K. \& {LaBonte}, B.~J. 2007, \apj, 671, 1034

\bibitem[{{Handy} {et~al.}(1999){Handy}, {Tarbell}, {Wolfson}, {Korendyke}, \&
  {Vourlidas}}]{1999SoPh..190..351H}
{Handy}, B.~N., {Tarbell}, T.~D., {Wolfson}, C.~J., {Korendyke}, C.~M., \&
  {Vourlidas}, A. 1999, \solphys, 190, 351

\bibitem[{{Heyvaerts} {et~al.}(1977){Heyvaerts}, {Priest}, \&
  {Rust}}]{1977ApJ...216..123H}
{Heyvaerts}, J., {Priest}, E.~R., \& {Rust}, D.~M. 1977, \apj, 216, 123

\bibitem[{{Hirose} {et~al.}(2001){Hirose}, {Uchida}, {Uemura}, {Yamaguchi}, \&
  {Cable}}]{2001ApJ...551..586H}
{Hirose}, S., {Uchida}, Y., {Uemura}, S., {Yamaguchi}, T., \& {Cable}, S.~B.
  2001, \apj, 551, 586

\bibitem[{{Howard} {et~al.}(1990){Howard}, {Harvey}, \&
  {Forgach}}]{1990SoPh..130..295H}
{Howard}, R.~F., {Harvey}, J.~W., \& {Forgach}, S. 1990, \solphys, 130, 295

\bibitem[{{Jing} {et~al.}(2004){Jing}, {Yurchyshyn}, {Yang}, {Xu}, \&
  {Wang}}]{2004ApJ...614.1054J}
{Jing}, J., {Yurchyshyn}, V.~B., {Yang}, G., {Xu}, Y., \& {Wang}, H. 2004,
  \apj, 614, 1054

\bibitem[{{Kim} {et~al.}(2006){Kim}, {Roh}, {Cho}, \&
  {Shin}}]{2006A&A...456..747K}
{Kim}, S.~S., {Roh}, H., {Cho}, K., \& {Shin}, J. 2006, \aap, 456, 747

\bibitem[{{Kopp} \& {Pneuman}(1976)}]{1976SoPh...50...85K}
{Kopp}, R.~A. \& {Pneuman}, G.~W. 1976, \solphys, 50, 85

\bibitem[{{Landi Degl'Innocenti}(1992)}]{1992soti.book...71L}
{Landi Degl'Innocenti}, E. 1992, {Magnetic field measurements}, ed. {Sanchez,
  F., Collados, M., \& Vazquez, M.}, 71--+

\bibitem[{{Liu} {et~al.}(2009){Liu}, {Lee}, {Karlick{\'y}}, {Prasad Choudhary},
  {Deng}, \& {Wang}}]{2009ApJ...703..757L}
{Liu}, C., {Lee}, J., {Karlick{\'y}}, M., {et~al.} 2009, \apj, 703, 757

\bibitem[{{Priest} \& {Forbes}(2000)}]{2000mare.book.....P}
{Priest}, E. \& {Forbes}, T. 2000, {Magnetic Reconnection}, ed. {Priest, E.~\&
  Forbes, T.}

\bibitem[{{Priest} \& {Forbes}(2002)}]{2002A&ARv..10..313P}
{Priest}, E.~R. \& {Forbes}, T.~G. 2002, \aapr, 10, 313

\bibitem[{{Priest} {et~al.}(1994){Priest}, {Titov}, {Vekestein}, \&
  {Rikard}}]{1994JGR....9921467P}
{Priest}, E.~R., {Titov}, V.~S., {Vekestein}, G.~E., \& {Rikard}, G.~J. 1994,
  \jgr, 99, 21467

\bibitem[{{Rubio da Costa} {et~al.}(2009){Rubio da Costa}, {Fletcher},
  {Labrosse}, \& {Zuccarello}}]{2009A&A...507.1005R}
{Rubio da Costa}, F., {Fletcher}, L., {Labrosse}, N., \& {Zuccarello}, F. 2009,
  \aap, 507, 1005

\bibitem[{{Schuck}(2005)}]{2005ApJ...632L..53S}
{Schuck}, P.~W. 2005, \apjl, 632, L53

\bibitem[{{Sterling} \& {Moore}(2005)}]{2005ApJ...630.1148S}
{Sterling}, A.~C. \& {Moore}, R.~L. 2005, \apj, 630, 1148

\bibitem[{{Wang} {et~al.}(2007){Wang}, {Liu}, {Jing}, \&
  {Yurchyshyn}}]{2007AAS...210.9321W}
{Wang}, H., {Liu}, C., {Jing}, J., \& {Yurchyshyn}, V. 2007, in Bulletin of the
  American Astronomical Society, Vol.~38, Bulletin of the American Astronomical
  Society, 214--+

\bibitem[{{Woodgate} {et~al.}(1981){Woodgate}, {Shine}, {Brandt}, {Chapman},
  {Michalitsianos}, {Kenny}, {Bruner}, {Rehse}, {Schoolman}, \&
  {Cheng}}]{1981ApJ...244L.133W}
{Woodgate}, B.~E., {Shine}, R.~A., {Brandt}, J.~C., {et~al.} 1981, \apjl, 244,
  L133

\bibitem[{{Zuccarello} {et~al.}(2009){Zuccarello}, {Romano}, {Farnik},
  {Karlicky}, {Contarino}, {Battiato}, {Guglielmino}, {Comparato}, \&
  {Ugarte-Urra}}]{2009A&A...493..629Z}
{Zuccarello}, F., {Romano}, P., {Farnik}, F., {et~al.} 2009, \aap, 493, 629

\end{thebibliography}

\end{document}